\documentclass[twocol]{ametsocV6.1}
\usepackage{booktabs}
\DeclareMathAlphabet{\mathpzc}{T1}{pzc}{m}{it}
\usepackage{multicol}
\usepackage{graphicx} %
\usepackage{lscape}
\usepackage{algorithm}
\usepackage{algpseudocode}
\usepackage[hidelinks]{hyperref}
\DeclareMathOperator{\evsym}{E}
\newcommand\ev[1]{\evsym\left\langle#1\right\rangle}
\newcommand{\wmode}{{\Psi}^\pm_\textrm{w}}

\title{Toward a Universal Framework for the Internal Gravity Wave Spectrum}
\authors{Leticia Fabre-Lima\aff{a}\correspondingauthor{Leticia Fabre-Lima, ldelima@umassd.edu}, Jeffrey Early\aff{b}, Miles A. Sundermeyer\aff{a}
}
\affiliation{\aff{a}{School for Marine Science and Technology University of Massachusetts Dartmouth, New Bedford, MA, USA}\\
\aff{b}{NorthWest Research Associates, Seattle, WA, USA}}
\abstract{The Garrett–Munk (GM) spectrum has long provided a canonical model of the oceanic internal gravity wave field. However, it relies on hydrostatic assumptions and idealized stratification that limit its applicability where non-hydrostatic dynamics, vertical boundary effects, or non-monotonic stratification are important. Here we develop a generalized framework for the internal wave spectrum based on non-hydrostatic vertical modes formulated in horizontal wavenumber–vertical mode space. Energetic orthogonality among wave modes requires that such a formulation be cast in horizontal wavenumber space rather than frequency space. In this formulation, the deformation radius associated with each vertical mode provides a proxy for distinguishing hydrostatic and non-hydrostatic regimes. Vertical modes are obtained numerically from the fixed-K Sturm–Liouville problem, allowing arbitrary stratification and multiple turning depths. Combined with a generalized spectral function, the formulation yields expected distributions of horizontal kinetic, vertical kinetic, and potential energy as functions of depth, frequency, and horizontal wavenumber. Example applications illustrate departures from GM theory associated with boundary effects and non-hydrostatic dynamics, including improved representation of vertical variance and high-frequency vertical kinetic energy, while reproducing observed features of horizontal wavenumber spectra.}

\begin{document}
\maketitle
\section{Introduction}
A major advance in describing the oceanic internal wave field was the development of the statistical and empirical internal wave model by \cite{GarrettMunk1972, GarrettMunk1975,GarrettMunk1979}, hereafter referred to collectively as GM. The GM spectrum synthesized available data at the time into a statistical representation of the depth-integrated energy distribution of the internal wave field, with internal wave modes expressed in terms of frequency and vertical mode number \citep{Levine_1983}. The model is formulated in two parts: (i) the solution of the internal-wave equation system defines the horizontal and vertical basis functions, and (ii) an empirical spectral function, derived from observations using a variety of instruments, prescribes the distribution of energy within that basis. A key feature of this approach is that the spectrum is independent of both stratification and observing platform. For a given stratification, the internal wave modes can be used to express the GM spectrum in terms of the horizontal and vertical spectra of the fluid velocity and buoyancy anomaly. The GM spectrum was not, at least initially, intended to provide all the answers; rather, it was proposed as a spectral model that captured the essential features of the internal wave field in the world ocean, and could therefore serve as a reference, or “strawman,” for comparing measurements and designing experiments \citep{Briscoe1975}.

Subsequent observational experiments revealed an unexpected degree of “universality” in the GM model; despite its simplicity, the spectrum appeared remarkably consistent in space and time \citep{Wunsch1976,Wunschwebb1979,Muller1978}.
This helped consolidate the view of GM as a baseline for the global ocean. However, deviations were also observed. Differences in the energy level and spectral shape were reported near topographic features, at the equator, and in regions of high mean shear \citep{Wunsch1976, Wunschwebb1979, Eriksen1985}. Surface forcing was found to further modify the spectrum, creating elevated energy under moderate winds, and anomalously high vertical coherence in the upper ocean \citep{Johnson1978, KatzBriscoe1979}.
\cite{polzin_2011} reviewed decades of observations and concluded that the GM spectral model fits wintertime conditions at Site D (a hydrographic section from Woods Hole to Bermuda) reasonably well,
but significant deviations occur elsewhere. Both regional and temporal variability were observed. In particular, many observations showed a stronger inertial peak than GM predicts.
Similarly, \cite{leBoyer_2021} examined variability of the internal wave field relative to GM using 2260 current meter records, finding that the observed continuum spectral energy $E_\text{cont}$ fell within $0.1 E_\text{GM} < E_\text{cont}< 5E_\text{GM}$. The observed spectral slope matched GM in the Atlantic Ocean, but was slightly steeper in the Pacific and Indian Oceans. Deviations were also observed near the seafloor in the ratio between observed energy and that predicted by GM. Many of these departures were not simply different spectral shapes, but pointed to dynamical processes not captured by GM, including boundary effects, variations in stratification, and non-hydrostatics.

One limitation of GM arises because it was designed for the deep ocean, where the buoyancy frequency varies gradually with depth and hence the Wentzel–Kramers–Brillouin-Jeffreys (WKBJ) approximation provides an accurate description of vertical wavenumber structure. This approximation breaks down where stratification varies strongly with depth; large departures are thus expected in the upper ocean \citep{KaseSiedler1980}, near the seafloor, and in the presence of turning depths. In addition, the GM spectrum essentially ignores vertical boundaries and is strictly hydrostatic. These shortcomings motivated several revisions and alternatives to GM. \cite{Desaubies_1973} derived non-hydrostatic WKBJ solutions for arbitrary stratification, obtaining a formulation that was more accurate \citep{Desaubies_1975}, but also considerably more complex to apply. \cite{Levine_2002} proposed a modified GM framework that retained much of the simplicity of the original model while incorporating vertical boundaries and turning depths through a modified WKBJ representation. Levine also noted that normalizing the GM frequency spectrum by the total integrated energy between $f$ and $N$ causes the predicted spectral level to scale with latitude and become unrealistically small near the equator. To address this issue, he introduced an alternative normalization referenced only to the high-frequency spectral level. He further showed that observations below the semidiurnal tidal frequency were generally “whiter” than predicted by GM, and suggested modifying the low-frequency dependence by introducing a different empirical spectral slope.

Despite decades of progress, the approximations underlying the variance relations and vertical structure functions commonly used in internal-wave spectral models do not appear to have been fully revisited. With advances in \textit{in situ} observations and numerical modeling, the oceanographic community has increasingly recognized the important role of non-hydrostatic internal waves in ocean dynamics, including energy transfer across the small scales at the end of the internal-wave spectrum toward turbulence. For example, \cite{Pinkel_2023} showed that such waves play an important role in balancing the large-scale forcing of the equatorial current system.
More broadly, finescale parameterizations of turbulent dissipation, which infer $\epsilon$ from shear and strain observations under the assumption of weakly nonlinear resonant interactions, depend explicitly or implicitly on GM theory as a reference, so that in regions where the internal wave spectrum departs from cannonical GM, estimates of $\epsilon$ may be biased \citep{Takahashi2021}.
These developments motivate the need for a non-hydrostatic internal gravity wave framework capable of representing the portion of the spectrum now accessible to modern observations and simulations.

\subsection{Scope and outline}
In this paper, we start by revisiting the linear solution for internal gravity waves, reviewing how different approaches to the vertical modes and vertical structure functions have evolved, and noting the shortcomings of each. To overcome previous limitations, we propose a new formulation that uses non-hydrostatic internal wave modes expressed in terms of horizontal wavenumber (as opposed to frequency) and vertical mode number to guarantee energetic orthogonality among wave modes. The proposed formulation accounts for non-hydrostatic dynamics, vertical boundary effects, and arbitrary stratification.

The paper is organized as follows. Section \ref{sec:linear-solution} reviews the well-known linear internal-wave solutions, highlights the statistical assumptions underlying an ensemble of waves, discusses hydrostatics and non-hydrostatics representations of the vertical structure proposed over time, and states the necessary conditions for formulating an orthogonal spectral model. Section \ref{sec:Results} presents the derivation of orthogonality conditions for internal-wave energy used to construct an energetically orthogonal vertical basis, introduces the proposed non-hydrostatic model, and provides example cases highlighting improvements over previous theories.
Section \ref{sec:discussion} highlights key implications of the revised horizontal wavenumber-vertical mode spectral formulation. Finally, Section \ref{sec:conclusions} summarizes and concludes.

\section{Linear internal wave solutions} \label{sec:linear-solution}
To facilitate comparison between different proposed representations of the internal gravity wave spectrum, we adopt a more modern notation than was originally used by \cite{GarrettMunk1972}, hereafter referred to as GM72. The governing linearized equations under the $f$-plane and Boussinesq approximations are:
\begin{subequations}
\label{eq:Sys}
    \begin{align}
        \label{eq:SysEQ1}
        \partial_t u- f\upsilon &=- \frac{1}{\rho_0} \partial_x p \\
        \label{eq:SysEQ2}
        \partial_t \upsilon+ fu &= - \frac{1}{\rho_0} \partial_y p \\
        \label{eq:SysEQ3}
        \partial_t w &=  -\frac{1}{\rho_0} \partial_z p - N^2(z) \eta\\
        \label{eq:SysEQ4}
        \partial_x u+ \partial_y \upsilon+ \partial_z w&=0 \\
        \label{eq:SysEQ5}
        \partial_t \eta - w &= 0
    \end{align}
\end{subequations}
\noindent where $p(x,y,z,t)$ and $\rho(x,y,z,t)$ are  pressure and density perturbation, respectively, defined such that total pressure $p_{\textnormal{tot}}(x,y,z,t)= p_0(z) +  p(x,y,z,t)$ and total density $\rho_{\textnormal{tot}}(x,y,z,t)= \Bar{\rho}(z) + \rho(x,y,z,t)$ where $\rho_0 \equiv\Bar{\rho}(0)$. The buoyancy frequency is defined as $N^2(z) \equiv -(g/\rho_0)\partial_z \bar{\rho}$. Throughout this manuscript we use a scaled density anomaly, which can be interpreted as the linear approximation to isopycnal displacement $\eta \equiv -\rho/\bar{\rho}_z$. We further assume periodic boundary conditions in the horizontal directions, $(x, y)$.

The simplified equations of motion \eqref{eq:Sys} admit several conserved quantities, including potential density, potential enstrophy, and total energy. The total energy is given by
\begin{equation}
\label{total-energy}
   E_0 \equiv \frac{1}{2 L_xL_y}  \int_{-D}^0 \int_A (u^2+\upsilon^2 + w^2 + N^2 \eta^2) \, dA \, dz,
\end{equation}
expressed here as a depth-integrated area-averaged quantity where $D$ represents full ocean water depth, and $A$ represents the horizontal domain \( [0, L_x] \times [0, L_y] \). When making the hydrostatic approximation, $\partial_t w$ is dropped from \eqref{eq:SysEQ3}, and as a consequence $w$ is dropped from \eqref{total-energy}.

\subsection{A single wave}
The system of equations given by \eqref{eq:Sys} has two types of solutions: two with non-zero frequency and one with zero frequency. The non-zero frequency wave solutions for the positive and negative frequency modes are given by
\begin{equation}
\wmode \equiv
    \begin{bmatrix}
          u_\pm\\ \upsilon_\pm \\ w_\pm\\ \eta_\pm \\ p_\pm\\
    \end{bmatrix} =
    \begin{bmatrix}
       \vspace{0.05in}   \frac{k\omega \mp f i \ell}{\omega K} F(z)  \\
       \vspace{0.05in}   \frac{\ell \omega \pm f i k}{\omega K} F(z) \\
       \vspace{0.05in}   - i K h \:G(z) \\
       \vspace{0.05in}   \mp\frac{ K h}{\omega}\;G(z)\\
       \vspace{0.05in}   \frac{\mp \rho_0 g K h}{\omega} F(z) \\
    \end{bmatrix} e^{i\theta_\pm}
    \label{eq:solution}
\end{equation}
\noindent where $K=\sqrt{k^2 + \ell^2}$ is the total horizontal wavenumber, $\theta_\pm=k x + \ell y \pm \omega t$, and functions $F(z)$ and $G(z)$ represent the vertical dependence of each mode with eigendepth $h$. Multiplying $\wmode$ by a complex amplitude $A$ and adding it to its complex conjugate produces a real-valued solution in $(x,y,z,t)$. The eigendepth, $h$, and the frequency, $\omega$, are related to the horizontal wavenumbers, $k$, $\ell$, through the dispersion relation,
\begin{equation}
  \omega = \sqrt{gh(k^2+\ell^2) + f^2}.
  \label{eq:DR}
\end{equation}
The terminology used here is that $\wmode$ is the \emph{wave mode}, which is comprised of spatial and temporal \emph{Fourier modes}, $e^{i(kx+\ell y)}$, $e^{i\omega t}$, and \emph{vertical modes}, $F(z)$, $G(z)$. Note that the $\pm$ indicates that there are two independent wave solutions at each $k,\ell,j$ (left and right propagating). The $\Psi^{-}_{w}$ mode is obtained from $\Psi^{+}_{w}$ by the substitution $\omega \mapsto -\omega$. The zero-frequency solution to \eqref{eq:Sys} is the geostrophic mode, sometimes referred to as the vortical mode, and is not relevant to the scope of the present paper.

The horizontal kinetic, vertical kinetic, and potential energies (HKE, VKE, and PE, respectively), averaged horizontally over a periodic domain for each individual wave mode solution per \eqref{eq:solution} are
\begin{subequations}{}
\label{eq:EnergyPieces}
      \begin{align}
          \label{eq:hke}
            \overline{\textrm{HKE}}(z) =& \frac{1}{2L_xL_y} \int_A(u^2 +\upsilon^2)\, dA %
            =\frac{A_\pm^2 h}{4} \left( 1+ \frac{f^2}{\omega^2} \right) \frac{1}{h} (F(z))^2, \\ %
            \label{eq:vke}
            \overline{\textrm{VKE}}(z) =& \frac{1}{2L_xL_y} \int_A w^2\, dA %
            =\frac{A_\pm^2 h}{4} \left( 1- \frac{f^2}{\omega^2} \right)  \frac{\omega^2}{g}  (G(z))^2, \\ %
            \label{eq:pe}
            \overline{\textrm{PE}}(z)=&\frac{1}{2L_xL_y} \int_A N^2\eta^2 \, dA%
            = \frac{A_\pm^2 h}{4}\left( 1- \frac{f^2}{\omega^2} \right)  \frac{N^2}{g} (G(z))^2.  %
      \end{align}
\end{subequations}
Here the energies are expressed in terms of $\omega$, but they can also be written in terms of $K$ using \eqref{eq:DR}. The depth-integrated sum of \eqref{eq:hke}-\eqref{eq:pe} is equivalent to \eqref{total-energy} and, when possible, we will normalize the vertical modes so that $E_0 = A_\pm^2 h/2$ for each individual wave.

\subsection{The eigenvalue problem}
Substitution of \eqref{eq:solution} into \eqref{eq:Sys} yields $\left(N^2(z) - \omega^2 \right) G=-g \partial_z F$ and  $F=h \partial_z G$ for the vertical momentum and mass continuity equations, respectively. Combined with the chosen constraints at the vertical boundaries, $\eta(0)=\eta(-D)=0$, these lead to a Sturm–Liouville problem that determines the vertical dependence of the system. This eigenvalue problem can be formulated in different ways. If we impose the hydrostatic approximation by neglecting the time variation of vertical velocity in \eqref{eq:SysEQ3}, the problem simplifies to
\begin{equation} \label{eq:HydroVertDependence} \partial_{zz} G + \frac{N^2(z)}{g h}G = 0. \end{equation}

\noindent From Sturm-Liouville theory, it follows that when $N^2(z)>0$, the eigenfunctions, $G(z)$, form a complete ordered basis for vertical modes $j=1...\infty$, where the vertical modes satisfy the orthogonality conditions,
\begin{subequations}
    \begin{align}
    \label{hydrostatic_G_eqn}
    \frac{1}{g} \int_{-D}^{0} N^2(z) G^i G^j , dz &= \delta_{ij}\\
    \label{hydrostatic_F_eqn}
    \frac{1}{h} \int_{-D}^0 F^i F^j , dz &= \delta_{ij}.
    \end{align}
\end{subequations}

\noindent Note that the functions $G(z)$ and $F(z)$ above are unique for each $j$, irrespective of frequency or horizontal wavenumber. Consequently, under the hydrostatic approximation, waves with different frequencies or horizontal wavenumbers may share the same vertical modes.

In contrast to the hydrostatic case, the non-hydrostatic case introduces additional dependence in frequency and horizontal wavenumber. The problem then needs to be solved in either the frequency domain (for a given $\omega$) or the horizontal wavenumber domain (for a given $K$). In the frequency domain, the governing equation for $G(z)$ becomes
\begin{equation}
    \label{eq:OmegaVertDependence}
    \partial_{zz} G_{\omega} + \frac{N^2(z) - \omega^2}{g h_{\omega}}G_{\omega} = 0,
\end{equation}

\noindent with corresponding orthogonality conditions
\begin{subequations}
    \begin{align}
    \label{eq:Omega_G_eqn}
    \frac{1}{g} \int_{-D}^{0} (N^2(z)-\omega^2) G^i_{\omega} G^j_{\omega} \, dz &= \delta_{ij}\\
    \label{eq:Omega_F_eqn}
    \frac{1}{h_{\omega}}\int_{-D}^0 F^i_{\omega} F^j_{\omega} \, dz &= \delta_{ij},
    \end{align}
\end{subequations}

\noindent applicable for two waves with the same frequency $\omega$. In our present notation, the subscript $\cdot_\omega$ indicates that the eignevalue problem is solved in the frequency domain.
Alternatively, in the horizontal wavenumber domain, the eigenvalue problem takes the form
\begin{equation}
    \label{eq: KVertDependence}
    \partial_{zz}G_K - K^2G_K = - \frac{N^2 - f^2}{g h_K},
\end{equation}

\noindent with corresponding orthogonality conditions
\begin{subequations}
    \begin{align}
    \label{eq:K_G_eqn}
    \frac{1}{g} \int_{-D}^{0} (N^2(z)-f^2) G^i_K G^j_K dz &= \delta_{ij}\\
    \label{eq:K_F_eqn}
    \frac{1}{h_K} \int_{-D}^0 \left( F^i_K F^j_K +  K^2 h^i_K h^j_K G^i_K G^j_K \right)  dz &= \delta_{ij},
    \end{align}
\end{subequations}

\noindent and with the subscript $\cdot_K$ now indicating that the eigenvalue problem is solved in the horizontal wavenumber domain.
\begin{figure*} 
    \centering  \includegraphics[width=12cm,angle=0]{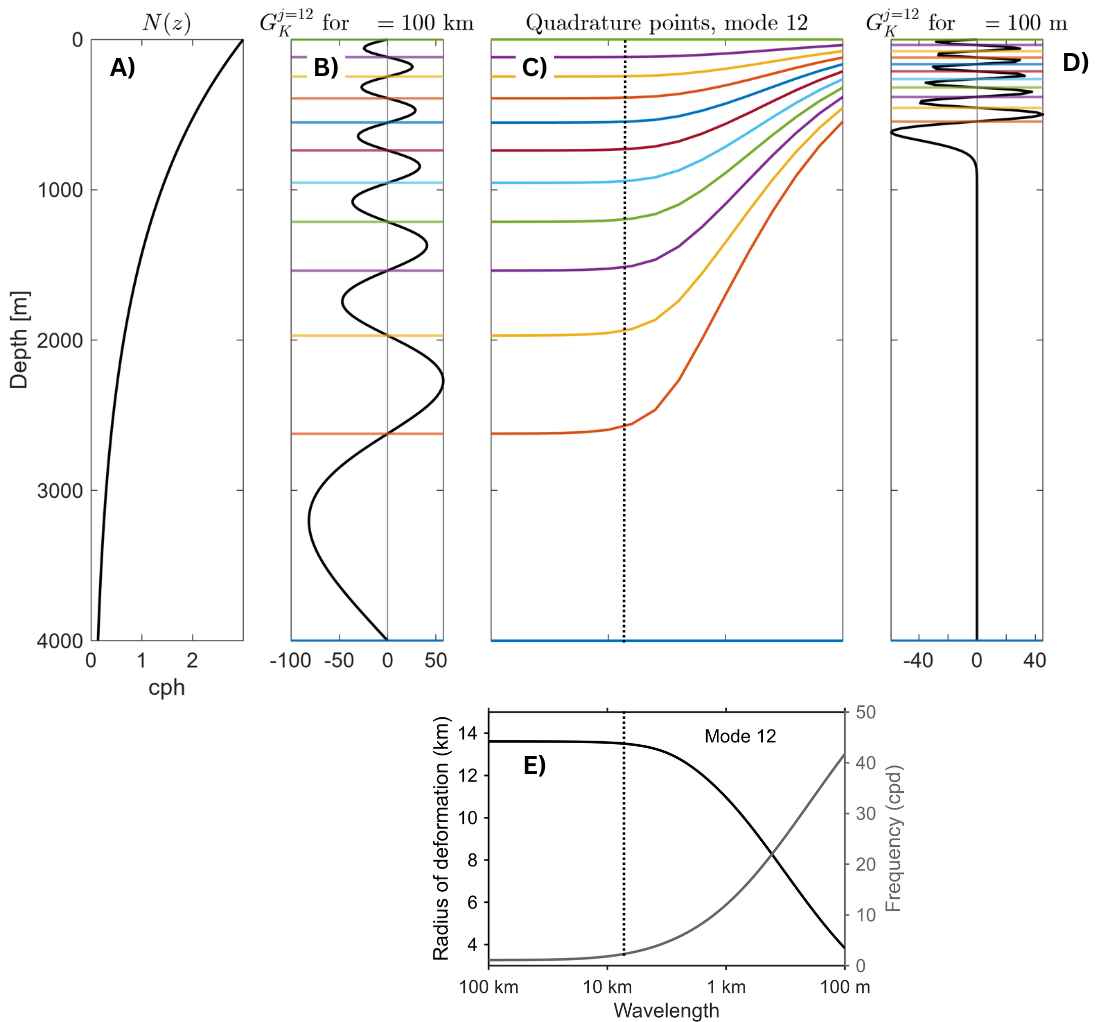}
    \caption{Top row first panel shows the stratification profile in terms of buoyancy frequency, $N(z)$. Second and fourth panels depict the vertical modes, $G^{j=12}_K(z)$, for waves ranging from 100 km to 100 m wavelengths. Third panel illustrates the horizontal wavelength-dependent variation in the depth of the quadrature points. Bottom panel shows the radius of deformation and frequency as a function of wavelength, with a dotted line indicating the approximate wavenumber where non-hydrostatic effects begin to become important.}
    \label{fig:quadPoints12}
\end{figure*}

An example illustrating the shape of the twelfth vertical mode, \( G^{j=12}_{K}(z) \), computed from solving \eqref{eq: KVertDependence} with exponential stratification, is shown in Figure \ref{fig:quadPoints12}. Per \eqref{eq:solution}, \( G^{j=12}_{K}(z) \) represents the vertical dependence of \( w \) and \( \eta \), computed using the stratification profile shown in the first panel of the figure. The second and fourth panels depict \( G^{j=12}_{K}(z) \) for waves with wavelengths of 100 km and 100 m, respectively. These panels highlight the variation in waveguide depth (determined by the respective turning depths) for the two different horizontal wavenumber, $K$, and hence frequency, $\omega$, waves. Key in this figure is that for the same vertical mode, a wave with a wavelength of 100 m has a higher frequency than one with a wavelength of 100 km. Consequently, the shorter-wavelength wave is confined to shallower depths, where its frequency is less than or equal to the local buoyancy frequency.

Meanwhile, in the third panel of Figure \ref{fig:quadPoints12}, the depths of the 12 zero-crossings of the mode are shown to depend on horizontal wavelength. Specifically, starting at wavelengths around 10 km, spacing between consecutive zero-crossings begins to decrease as wavelength shortens. This behavior reflects the onset of non-hydrostatic dynamics, where the vertical modes become $K$-dependent, in contrast to the hydrostatic case discussed above.  Finally, the variation in the waveguide depth is reflected in the bottom panel, which depicts the vertical mode deformation wavelength, and which is proportional to the eigenvalue $2\pi (g h_{\omega})^{1/2}/f$ in \eqref{eq: KVertDependence}.

\subsection{An ensemble of waves}
\label{subsec:ensemble}
The energy expressions in \eqref{eq:EnergyPieces} are for a single internal wave solution with a specific vertical mode and horizontal wavenumber. However, the internal gravity wave field as observed in the ocean is conceptually an ensemble of waves, i.e.,
\begin{equation}
\label{eq:EnsembleWave}
    \begin{bmatrix}
          u(x,y,z,t) \\
          \upsilon(x,y,z,t) \\
          w(x,y,z,t) \\
          \eta(x,y,z,t)\\
          p(x,y,z,t) \\
    \end{bmatrix} = \sum_\textrm{all waves}  A_\pm(\cdot)  \wmode + \textrm{c.c.}
\end{equation}
where the complex wave amplitudes $A_\pm(\cdot)$ may be a function of $(k,\ell,j)$ or $(\omega,j)$ and ``c.c." represents complex conjugate terms, making the total field real-valued.

An \textit{energy spectrum} specifies how total energy is distributed in wave mode space, i.e., the spectral domain. To formulate such a spectrum, it is necessary to ensure energy conservation when transforming between physical space and the spectral domain. This is achieved through depth-integrated energy orthogonality of the hydrostatic wave modes,
\begin{equation}
\label{eq:GMTheoreticalSpec}
\frac{1}{A}
\int_{-D}^0 \int_A
\left( u^2 + \upsilon^2 + N^2 \eta^2 \right)
\, dA \, dz = \sum_\textrm{all waves}  \frac{h^j}{2} A^2_\pm(\omega,j).
\end{equation}
\noindent
 Equation \eqref{eq:GMTheoreticalSpec} is analogous to Parseval’s theorem, where in this case the left-hand side defines the spatial distribution of energy, while the right-hand defines its distribution in the spectral domain, and thus $h^j A^2_\pm(\omega,j) / 2$ defines the `energy spectrum'.
 
 The energy spectrum can be computed exactly in a model where the full fluid state is known.
 However, to obtain a quantity that can be compared with measurements, in practice this spectrum must be combined with a statistical description of the wave field. A key assumption in this formulation (and elsewhere) is that the wave field is a stationary stochastic process where the amplitude of each wave, $A_\pm$, is assumed to be an uncorrelated Gaussian random variable with zero mean.\footnote{As it is complex-valued it is actually two random variables, generating an amplitude and phase.} A stationary wave field can be formulated through expectation, where we use $\ev{X}$ to denote the expectation of random variable $X$. Assuming we are at a mooring where $x=x_0$, $y=y_0$, $t=t_0$ then $\ev{u + i \upsilon}(z)=0$, $\ev{w}(z)=0$ and $\ev{\eta}(z)=0$ follows immediately from \eqref{eq:EnsembleWave} because the wave amplitudes are assumed to be zero-mean.

 In the GM72 spectrum, the \emph{expected value} of the \emph{energy spectrum} is taken to be
 \begin{equation}
 \label{eq:GMspectrum-omega}
 \ev{h^j A^2(\omega,j)} = E_0 B(\omega) H(j),
 \end{equation}
where $\pm$ waves are assumed to be statistically identical. \(B(\omega)\) and \(H(j)\)
are the well-known empirical functions describing the frequency and vertical-mode dependence of the spectrum, for which there are several variations in the literature, and $E_0$ is the reference energy level with units m$^3$s$^{-2}$. The reference energy is often defined as $E_0 = E_\textrm{GM} b^3 N_0^2$, where $E_\textrm{GM} =6.5 \cdot 10^{-5}$ is the non-dimensional magnitude, $b = 1300$ m is the e-folding scale of the pycnocline, and $N_0=5.2 \cdot 10^{-3}$ rad s$^{-1}$ is the maximum buoyancy frequency. With an expected amplitude of each wave mode that compose an ensemble of waves, \eqref{eq:EnsembleWave} can be combined with \eqref{eq:GMTheoreticalSpec} to show
\begin{subequations}
\label{eq:EK-uv-eta}
    \begin{align}
        \ev{u^2 + \upsilon^2}(\omega,z) &=  \frac{E_0}{b} B(\omega) \left( 1+ \frac{f^2}{\omega^2}  \right) \Phi(z) \\
        \ev{\eta^2}(\omega, z) &= \frac{E_0}{N_0^2 b} B(\omega) \left( 1 -\frac{f^2}{\omega^2}  \right) \Gamma(z),
    \end{align}
\end{subequations}
the so-called `moored spectrum' \citep{Levine_2002}. Note that the cross-terms, i.e., nonlinear contributions involving different individual wave modes, vanished because $\ev{X_i X_j} = 0$ for uncorrelated random variables. Additionally, the contribution of individual vertical modes is represented through vertical structure functions defined by their weighted-square values, $\Phi(z) = \sum (b/h^j) H(j) (F^j)^2$ and $\Gamma(z) = (N_0^2 b /g ) \sum H(j) (G^j)^2$, where the scales $b$ and $N_0$ are used to make them unitless.

Under the above approximation, expected values of horizontally averaged horizontal velocity, vertical velocity, and isopycnal displacement (GM72 (4.5), equivalent to \eqref{eq:EK-uv-eta}) have the following vertical structure functions
\begin{subequations}
\label{vertical-structure-gm}
\begin{align}
\label{vertical-structure-gm-phi}
\Phi_{\text{GM}}(z) &= \frac{N(z)}{N_0},\\ \label{vertical-structure-gm-gamma}
\Gamma_{\text{GM}}(z) &= \frac{N_0}{N(z)}.
\end{align}
\end{subequations}

A point worth emphasizing is the distinction between \eqref{eq:GMTheoreticalSpec} and \eqref{eq:EK-uv-eta}, which is often blurred in the literature. Equation \eqref{eq:GMTheoreticalSpec} expresses how energy in a three-dimensional stratified fluid is distributed in terms of wave modes at a given instant in time. Quadratic cross-terms vanish due to energy orthogonality. The moored spectrum \eqref{eq:EK-uv-eta} follows from \eqref{eq:GMTheoreticalSpec}, but only after assuming a statistical distribution of wave energy and restricting to a fixed spatial location. The quadratic cross-terms that would otherwise appear here vanish under expectation, given the assumption that the individual wave amplitudes are uncorrelated. The two expressions are thus very distinct concepts.

\subsection{The approximations of GM72}
Using the above framework, the full theory in GM72 includes an approximation of the vertical structure functions, $\Phi$ and $\Gamma$, derived from the vertical eigenvalue problem. Specifically, the system of equations in GM72 (1.1) is solved under the assumption of a separable solution in horizontal and vertical coordinates. The Sturm-Liouville problem is posed in the frequency domain, per GM72 (1.3). Initially, the vertical domain was bounded between the surface and local depth, $-D$, with boundary conditions $G(0) = G(-D) = 0$. Note that GM72 (1.1) and GM72 (1.3) are equivalent to our \eqref{eq:Sys} and \eqref{eq:OmegaVertDependence}, respectively.

Another major approximation in GM72 was to assume exponential stratification, $N(z) = N_0 e^{z/b}$, where \( N_0 \) is a reference buoyancy frequency. Under this assumption, GM72 (1.3) admits solutions for the vertical modes in terms of Bessel functions of the first and second kinds, \( J_\nu \) and \( Y_\nu \), with order \( \nu \) determined by the stratification and horizontal wavenumber (GM72 equation (1.10)). For large wavenumbers, these solutions reduce to asymptotic forms involving Airy functions (GM72 equation (1.18)). Alternatively, this result can be derived more straightforwardly using the WKBJ-like approximation, as noted by a reviewer in GM72 and demonstrated in subsequent works \citep{Desaubies_1973,Levine_2002}.

A final major approximation in GM72 was the neglect of vertical boundaries, restricting the solution to the ocean interior and to depths away from the turning depth for a given frequency. This necessarily implied that waves are hydrostatic. Vertical modes are replaced by their mean-square values, defined as averages over vertical modes above the turning depth and set to zero below it (GM72 equation (2.10)).

In what follows we compare subsequent historical formulations of the internal wave spectrum in a common notation before presenting the results in \S\ref{sec:Results}. In all cases (with one notable exception), total energy is assumed to be a separable function of frequency and vertical mode, with the negative and positive waves assumed to be statistically equivalent.

\subsection{Extensions to non-hydrostatics and arbitrary stratification} \label{Sec:TheoreticalBackground}
Following the publication of GM72, a few studies have proposed modifications to the vertical basis functions to account for non-hydrostatic effects, or to construct a fully non-hydrostatic spectrum. In particular, \cite{Desaubies_1973} derived a uniformly non-hydrostatic solution using a WKBJ approach, where vertical mode solutions for arbitrary stratification in a bounded domain were obtained and expressed in terms of Airy functions (equations (2.11) and (2.12) of \cite{Desaubies_1973}). The form of the solution depended on the presence or absence of a turning depth. When it was absent, the solution was sinusoidal from the surface to the bottom. When it was present, vertical modes could be approximated as sinusoidal above the turning depth and exponentially decaying below it (equation (2.13) of \cite{Desaubies_1973}).

Although the \cite{Desaubies_1973} solution allowed for the presence of a turning depth, it admitted at most one. This required the stratification to be monotonic. If $N(z)$ is non-monotonic, the condition $N(z)=\omega$ could be satisfied at multiple depths, producing additional turning depths. However, for nonmonotonic stratification, the solution provided only a local approximation. When exponential stratification is assumed, the WKBJ solution of \cite{Desaubies_1973} closely resembles that of GM72, indicating that the former represents a generalization of the latter.

Away from turning points, \cite{Desaubies_1973} derived the vertical structures, \(\Phi_{\text{WKBJ}}(z)\) and \(\Gamma_{\text{WKBJ}}(z)\), by averaging many vertical modes, giving
\begin{subequations}
\label{eq:PhiGammaDes}
    \begin{align}
        \label{eq: PhiDes}
        \Phi_{\text{WKBJ}}(z) &= I^{-1}\frac{({N(z)^2 -\omega^2})^{1/2}}{N_0},\\
        \label{eq: GammaDes}
        \Gamma_{\text{WKBJ}}(z) &=  I^{-1}\frac{N_0}{(N(z)^2 -\omega^2)^{(1/2)}},
    \end{align}
\end{subequations}
\noindent where,
\begin{equation}
    I = \int_0^D \frac{(N^2 - f^2)}{(N^2 - \omega^2)^{1/2}} \, dz.
\end{equation}
\noindent Near the turning depth \cite{Desaubies_1973} presented a discrete solution for $\Phi_{\text{WKBJ}}(z_T)$ and $\Gamma_{\text{WKBJ}}(z_T)$ by computing the convergence in the limit \( z \to z_T \).

To the best of our knowledge, \cite{flatte_1979}, in a monograph to which Walter Munk contributed, was the first to formulate and solve the internal wave problem in the horizontal wavenumber domain. In their approach, the eigenvalue problem was formulated for a given horizontal wavenumber (equation (3.2.16) of \cite{flatte_1979}, corresponding to \eqref{eq: KVertDependence} in the present paper). This represented an important contribution, as it was not clear \textit{a priori} whether a formulation in the horizontal wavenumber domain or in the frequency domain is more suitable for deriving a spectral model for non-hydrostatic waves in non-constant stratification. This issue is discussed further in \text{\S}\ref{sec:Results}\ref{Sec:Orthogonality}.

A different modified spectral formulation was proposed by \cite{Levine_2002} to address the treatment of vertical boundaries and turning points within the GM framework. While deriving a solution with vertical boundaries, he retained the frequency-domain formulation and the analytical advantages of the WKBJ approximation. Following the eigenfunctions in Levine's work (his equations (15) and (16)), the vertical structure functions in his approach, $\Phi^\textrm{L}$ and $\Gamma^\textrm{L}$, are
\begin{subequations}
    \begin{align}
                \label{eq: GammaLev}
           \Phi^{\text{L}}(z) &= \sum_j b^2 \alpha_j^2 \frac{2N(z)}{N_0} \cos^2\left( j\pi \frac{\xi(z)}{D(\omega)}\right) \\
           \label{eq: PhiLev}
            \Gamma^{\text{L}}(z) &= \sum_j \frac{2(\omega^2 -f^2)}{N(z)N_0} \sin^2\left( j\pi \frac{\xi(z)}{D(\omega)}\right),
    \end{align}
\end{subequations}
\noindent where $\alpha_j$ is the eigenvalue and the stretched vertical coordinate was given by integral starting from the turning depth
\begin{equation}
       \xi(z) = \frac{1}{N_0} \int_{z_T}^z N(z)\,dz.
\end{equation}
The maximum value of $\xi$ was defined to be $D(\omega) \equiv \xi(0)$, which is the waveguide thickness (in stretched units) at frequency $\omega$.
Again, there was a challenge near turning depths; namely, the resulting vertical modes truncated energy below the turning depth, introducing an additional source of error. Particularly at higher frequencies, the WKBJ modes exhibit their largest deviations from the exact solution around the turning depth. This notwithstanding, an advantage of the formulation of \cite{Levine_2002} was that it retained the effect of layer thickness associated with turning depths in non-hydrostatic dynamics, while still maintaining a simple analytical form.

Summarizing work to date, in its original form, the Garrett–Munk model was intended as a foundational framework for describing the distribution of internal wave energy, and to guide experimental design, rather than as a complete representation of the internal wave field \citep[e.g.,][]{Briscoe1975}. Later formulations addressed certain simplifying assumptions, but many retained key limitations to preserve analytical simplicity. Table~\ref{tab:comparison} summarizes key aspects of the studies reviewed in this section and contrasts them with the improvements proposed here. With more than 50 years of accumulated knowledge and computational advances, it is now possible to develop a non-hydrostatic model that incorporates vertical boundary conditions, allows for arbitrary stratification, accommodates multiple turning depths, and provides a full solution to the eigenvalue problem without sacrificing practical usability.

\begin{table*}[t]
    \caption{Comparison of assumptions and methods across previous and current studies.}
    \centering
    \small
    \begin{tabular}{lccccc}
        \hline\hline
        & GM72 & \cite{Desaubies_1973} & \cite{flatte_1979} & \cite{Levine_2002} & This paper \\
        \hline
        Stratification & Exponential & Arbitrary monotonic & Arbitrary monotonic & Arbitrary monotonic & Arbitrary \\
        Basis domain & $(k,\ell, \omega)$ &  $(k,\ell, \omega)$ &  $(k,\ell,j)$ &  $(k,\ell, \omega)$ & $(k,\ell,j)$ \\
        Vertical boundaries & Not bounded & Mixed(?) & Bounded & Bounded & Bounded \\
        Hydrostatic? & Yes & No & No & Yes & No \\
        Solution method & Bessel + Airy & WKBJ + Airy & - & WKBJ + Airy & Computational (Chebyshev) \\
        Turning points? & No & One & - & One & Multiple \\
        \hline
    \end{tabular}
    \label{tab:comparison}
\end{table*}

\section{Results}
\label{sec:Results}
This section develops the non-hydrostatic internal-wave spectral framework in four steps. First, \S\ref{sec:Results}\ref{Sec:Orthogonality} establishes the orthogonality conditions required for defining non-hydrostatic spectrum. Second, \S\ref{sec:Results}\ref{Sec:SpectralFun} introduces a generalized spectral function formulated in horizontal wavenumber–vertical mode space. Third, \S\ref{sec:Results}\ref{sec:NonHydroSpectral} describes how the non-hydrostatic spectral model is constructed by combining the numerical vertical modes, orthogonality conditions, and generalized spectral function into expected energy distributions for the internal-wave field. Finally, \S\ref{sec:Results}\ref{sec:EnergyDistr} examines the resulting energy distributions through example applications.

\subsection{Orthogonality of internal wave energy} \label{Sec:Orthogonality}
In \S\ref{sec:linear-solution}, we reviewed how a wave spectrum may be formulated in either the frequency or the horizontal wavenumber domain. In the hydrostatic limit, both approaches produce orthogonal wave modes and satisfy energy orthogonality \eqref{eq:GMTheoreticalSpec}. However, this equivalence does not automatically hold for non-hydrostatic waves. Specifically, the following analysis proves that to maintain orthogonality, and thus to define an energy spectrum, the internal wave spectrum must be formulated in the wavenumber domain.
The full derivation is given in Appendix \ref{A:Othogonality}, while a brief summary is presented below.

The vertical modes obtained from  $\omega$-eigenvalue problem \eqref{eq:OmegaVertDependence} or the $K$-eigenvalue problem \eqref{eq: KVertDependence} are orthogonal by Sturm–Liouville theory, as previously noted. However, this property alone is not sufficient for energy orthogonality of the wave modes \eqref{eq:solution} with respect to the non-hydrostatic energy \eqref{total-energy}. Orthogonality must apply to the full ensemble of wave modes \eqref{eq:EnsembleWave}, not just to the vertical eigenfunctions. The following linear operation must therefore be satisfied
\begin{align}
\label{eq:linearOperation}
E &= \frac{1}{L} \int_{-D}^0 \int_A
\left( |\mathbf{u}_a|^2 + |\mathbf{u}_b|^2 + N^2(\eta_a^2 + \eta_b^2) \right)\, dA\, dz \\
&= \frac{1}{L} \int_{-D}^0 \int_A
\left( |\mathbf{u}_a+\mathbf{u}_b|^2 + N^2(\eta_a+\eta_b)^2 \right)\, dA\, dz . \nonumber
\end{align}
\noindent where $\mathbf{u}=(u,v,w)$ and $|\mathbf{u}|^2 = u^2+v^2+w^2$.
In other words, the energy of a summed wave field must equal the sum of the energies of its individual components. This property allows total energy to be represented as a distribution over waves with distinct attributes (e.g., mode, wavenumber, and frequency). Without energetic orthogonality, the partitioned energies cannot be summed to recover the correct total energy \cite[e.g.,][]{kelly2016}.

Appendix~\ref{A:Othogonality} considers energy orthogonality of the wave modes with vertical modes from the $K$ and $\omega$-eigenvalue problems. For the $K$-eigenvalue problem (or fixed-time), consider the case where two waves given by \eqref{eq:solution} have the same wavenumber but different vertical modes. Evaluating the linear operation \eqref{eq:linearOperation}, the cross terms assume a form that enables application of the orthogonality conditions \eqref{eq:K_G_eqn} and \eqref{eq:K_F_eqn}, whereby they vanish.
However, the same does not hold for the $\omega$-eigenvalue problem (fixed-location).
Consider two waves with the same frequency but different horizontal wavenumbers and vertical modes. Substituting these into \eqref{eq:linearOperation} and applying orthogonality conditions \eqref{eq:Omega_G_eqn} and \eqref{eq:Omega_F_eqn} the cross terms in horizontal kinetic energy vanish, but persist in the vertical kinetic and potential energies.
Key here is that in the $\omega$-eigenvalue problem the orthogonality conditions do not eliminate all cross terms, but they do in the $K$-eigenvalue problem. Internal wave modes formulated in the frequency domain are only orthogonal for hydrostatic waves.
The orthogonality condition for non-hydrostatic internal gravity wave is thus given by
\begin{equation}
\label{eq:NHSTheoreticalSpec}
\frac{1}{A}
\int_{-D}^0 \int_A
\left( |\mathbf{u}|^2 + N^2 \eta^2 \right)
\, dA \, dz = \sum_\textrm{all waves}  \frac{h^j}{2} A^2_\pm(k,\ell,j),
\end{equation}
which defines the energy spectrum for non-hydrostatic waves, just as \eqref{eq:GMTheoreticalSpec} did for hydrostatic waves.  Notable in \eqref{eq:NHSTheoreticalSpec} compared to \eqref{eq:GMTheoreticalSpec} is that vertical velocity is now included in the kinetic energy term on the lhs, and the wave amplitudes on the rhs depend on horizontal wavenumber and vertical mode rather than frequency and vertical mode.

Although the non-hydrostatic spectrum satisfying orthogonality must be formulated in horizontal wavenumber space, it can be mapped to the frequency domain via the dispersion relation. Each wave characterized by $(K,j)$ is associated with a unique frequency $\omega(K,j)$, allowing the spectral energy to be accumulated into frequency bins.
By analogy with GM, the expected non-hydrostatic IGW spectrum is defined as
\begin{equation}
 \label{eq:GMspectrum-K}
 \ev{h^j A^2(k,\ell,j)} = E_0 S(K,j)
 \end{equation}
where $\pm$ waves are assumed to be statistically identical.

\subsection{Spectral function} \label{Sec:SpectralFun}
Thus far, our analysis has focused on the first part of defining the spectrum: solving the linear system in ($\ref{eq:Sys}$) to obtain a set of orthogonal horizontal and vertical bases. The second step in defining the spectrum is to specify how energy is distributed across these bases, which is done empirically using a spectral function derived from observational data.

Garrett and Munk showed in a series of studies \citep{GarrettMunk1972, GarrettMunk1975, GarrettMunk1979} how measurements from various instruments could be synthesized into the expected internal wave spectrum \eqref{eq:GMspectrum-omega}. Here we begin by adopting the formulation of \citet{Munk_1981}, in which the spectral functions \( B(\omega) \) and \( H(j) \) are defined as
\begin{subequations}
    \begin{align}
        \label{eq:B}
        B(\omega) = \frac{f}{\omega} \frac{B_0}{\sqrt{ \omega^2 - f^2}},\\
        \label{eq:M}
        H(j)= \frac{H_0}{(j^2 + j_\ast^2)^{5/4}}.
    \end{align}
    \label{eq:B,M}
\end{subequations}

As shown in \text{\S}\ref{sec:Results}.\ref{Sec:Orthogonality}., energetic orthogonality for the non-hydrostatic internal wave spectrum requires formulation in the horizontal wavenumber domain. We therefore modify the spectral function \eqref{eq:B} by expressing it in terms of horizontal wavenumber \( K \) through the dispersion relation \eqref{eq:DR}, while preserving total variance,
\begin{equation}
    \int_K \sum_j S(K,j) \, dK = \int_\omega \sum_j B(\omega) H(j) \, d\omega.
\end{equation}
This yields a spectral function \( S(K,j) \) formulated as
\begin{equation}
    S(K,j) = \frac{L_{K}^{j}S_0}{\left((L_{K}^{j})^2 K^2 + 1\right)(j^2 + j_\ast^2)^{5/4}}.
    \label{eq:S1}
\end{equation}
The power dependence of 5/4 in the denominator and the parameter $j^\ast$, typically set to $3$, together control the rate at which energy decreases with increasing vertical mode number. The normalization constant $S_0$ is chosen such that the combined summation over modes and integration over wavenumber equals unity by construction, analogous to the constants $B_0$ and $H_0$ in the GM formulation.

In the hydrostatic GM formulation, the radius of deformation for vertical mode $j$ is independent of either horizontal wavenumber or frequency, and is given by
$L^j = (N_0 b)/(fj\pi)$ [see Appendix~\ref{A:ApproxGM}]. In the non-hydrostatic case, the radius of deformation for a given mode $j$ depends on $K$ and is given by
\begin{equation}
\label{eq:radDefS}
    L_{K}^{j} = \sqrt{g h_{K}^{j}}/f.
\end{equation}
 Despite this added dependence, the inverse relationship between vertical mode number and radius of deformation is retained (via the dependence of $h^j_K$ on $j$). A more general form of \eqref{eq:S1} follows from substituting $j$ in \eqref{eq:S1} by the proportional relation $L_K^j \propto 1/j$:
\begin{equation}
    \label{eq:generalSpec}
    S(K,j) \equiv \frac{L_{K}^{j} S_0}{\left((K/k_{\ast})^2 + 1\right)^{m_k}\left((L_{\ast}/
    L_{K}^{j})^2 +1 \right)^{m_j}},
\end{equation}
\noindent where $k_\ast = 1/L_{K}^{j}$ and $L_\ast = 1/L^{j^\ast}_K$ denote the horizontal wavenumber and horizontal wavelength roll-off scales, respectively. Both quantities still depend on $K$, but this dependence is suppressed in the notation for brevity. The parameters $m_j$ and $m_k$ are slope exponents that provide flexibility in spectral fitting. A key aspect of this formulation is that for non-hydrostatic internal waves, the dependence on horizontal wavenumber and vertical mode is not strictly separable; consequently, their spectral slopes cannot be controlled independently, in contrast to the GM spectrum formulated in the frequency domain as per \eqref{eq:B,M}.

\subsection{Non-hydrostatic internal gravity waves spectral model} \label{sec:NonHydroSpectral}
Non-hydrostatic vertical bases are obtained by numerically solving \eqref{eq: KVertDependence} in stretched coordinates and projecting the solution onto Chebyshev polynomials, a method that provides good accuracy even for high vertical modes \citep{Early_2020}. Here the strong dependence of vertical modes and their radius of deformation on horizontal wavelength must be considered when choosing how to represent the vertical dependence of solutions to \eqref{eq:Sys}. Any formulation of the Sturm–Liouville problem in which the vertical dependence is independent of both frequency or horizontal wavenumber is necessarily hydrostatic. In such a case, waves with very different horizontal scales—for example, 100 km and 100 m—are assigned identical vertical modes, leading to a misrepresentation of internal wave energy (Figure \ref{fig:quadPoints12}).

\begin{figure}
    \centering
    \includegraphics[width=8cm]{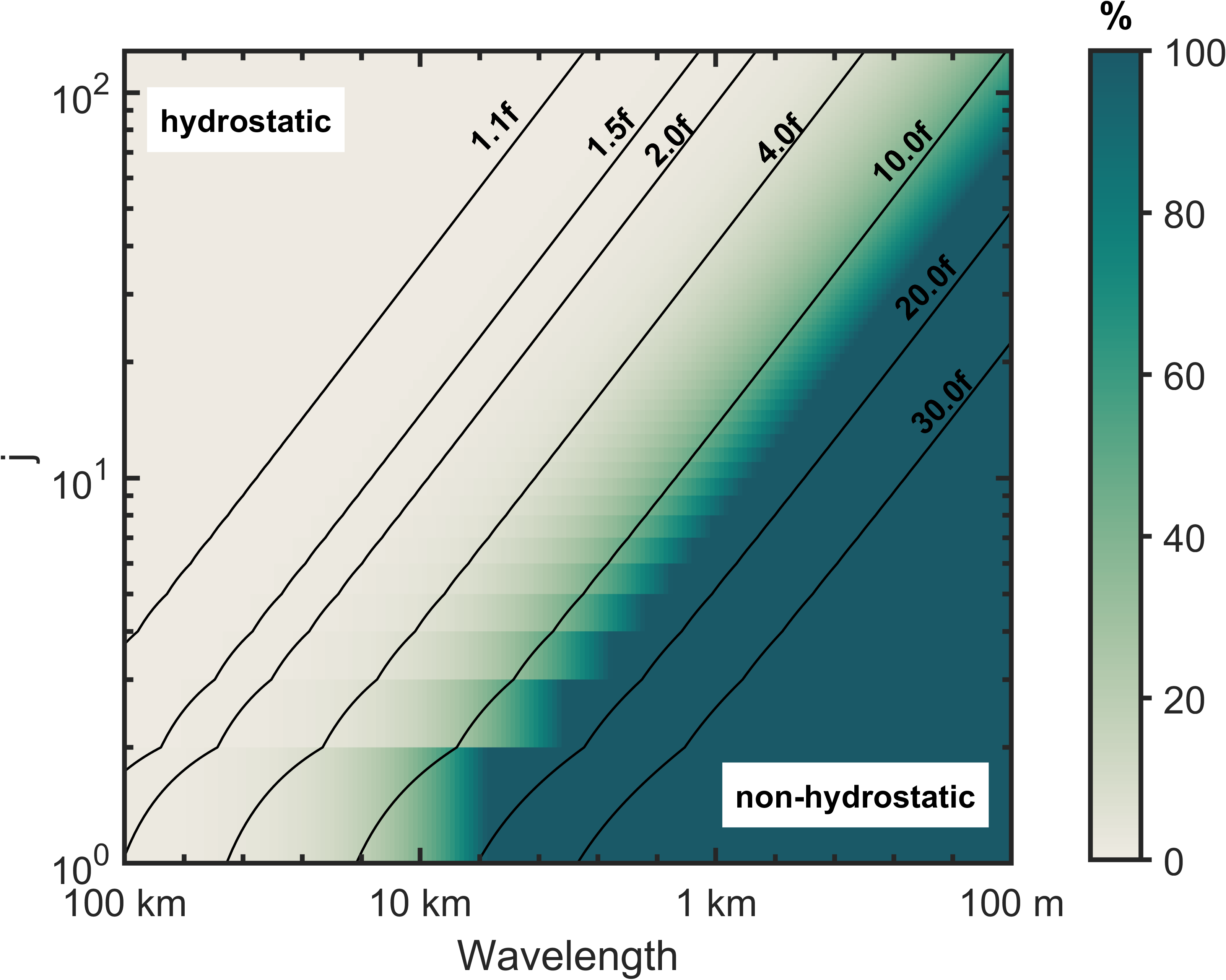}
    \caption{Difference between the non-hydrostatic radius of deformation, $L^j_K$, and the hydrostatic limit, $L^j_{K \to \infty}$, expressed as a percentage of $L^j_{K \to \infty}$, for the case of exponential stratification, shown as a function of horizontal wavelength $K$ and vertical mode number $j$. Lighter shading indicates small differences and darker shading larger differences. Note, the color scale saturates at 100\%. Black contours denote normalized frequency $\omega/f$. The upper-left region corresponds to the hydrostatic regime, and the lower-right to the non-hydrostatic regime. }
    \label{fig:criterion}
\end{figure}

Figure \ref{fig:criterion} shows the difference between the non-hydrostatic radius of deformation and its hydrostatic limit, expressed as a percentage of the hydrostatic limit, for the case of exponential stratification. This difference provides a practical measure of the error incurred when the hydrostatic approximation is applied outside its range of validity, and serves as a proxy for distinguishing hydrostatic and non-hydrostatic regimes. Overplotted in the figure are contours of normalized frequency, $\omega/f$, as a function of mode number and horizontal wavenumber. Because the deformation radius is linked to the dispersion relation, contours of constant $\omega/f$ align approximately with values of percent difference: $\omega = 4f$  corresponds to approximately 10\% difference, and $\omega= 10f$ to approximately 50\%. Note that while the color scale saturates at 100\%, values in the lower right region of the figure (corresponding to waves with frequencies exceeding $20f$) exceed that value.

Our approach to developing a non-hydrostatic internal wave spectral model introduces three major improvements relative to previous formulations. First, energy is represented through the whole water column and is not forced to vanish below the turning depth, as in \cite{Levine_2002}. This is physically important because internal waves do have some energy below their turning depth. Second, the present formulation accommodates nonmonotonic stratification, which is not supported by any of the solutions reviewed in \S\ref{sec:linear-solution}\ref{Sec:TheoreticalBackground}. Third, although \cite{Desaubies_1973} obtained a piecewise continuous analytical solution for non-hydrostatic internal waves, his solution is constructed by treating three regions separately (above, near, and below the turning depth) and then matching them. By contrast, the numerical approach presented here solves the problem continuously over the full vertical domain.

The preceding sections have established the linear solution  \eqref{eq:solution} for an ensemble of internal gravity waves \eqref{eq:EnsembleWave}, the orthogonality of the wave modes (\S\ref{sec:Results}\ref{Sec:Orthogonality}), and a generalized empirical spectral function \eqref{eq:generalSpec}. These elements are now combined to construct the non-hydrostatic moored spectrum, in contrast to the hydrostatic spectrum \eqref{eq:EK-uv-eta}. Here total energy is written as $E_0 S(K,j)$, where $E_0$ is again the `GM reference' level with units m$^3$s$^{-2}$, and $S(K,j)$ is given by \eqref{eq:generalSpec} with  $\int_K \sum_j S(K,j)= 1$ as per previous discussions. The expected values of velocity and vertical displacement are then given by
\begin{subequations}
\label{eq:EK-gen}

\begin{equation}
\label{eq:EK-uv-gen}
\begin{aligned}
\ev{u^2+\upsilon^2}(K,j,z)
&= \int_K \sum_j \frac{E_0}{b} S(K,j) \\
&\quad \cdot
\left(1+\frac{f^2}{(\omega_K^j)^2}\right)
\left(F_K^j(z)\right)^2\,dK,
\end{aligned}
\end{equation}

\begin{equation}
\label{eq:EK-w-gen}
\begin{aligned}
\ev{w^2}(K,j,z)
&= \int_K \sum_j \frac{E_0}{N_0^2 b} S(K,j) \\
&\quad \cdot
\left((\omega_K^j)^2-f^2\right)
\left(G_K^j(z)\right)^2\,dK,
\end{aligned}
\end{equation}

\begin{equation}
\label{eq:EK-ets-gen}
\begin{aligned}
\ev{\eta^2}(K,j,z)
&= \int_K \sum_j \frac{E_0}{N_0^2 b} S(K,j) \\
&\quad \cdot
\left(1-\frac{f^2}{(\omega_K^j)^2}\right)
\left(G_K^j(z)\right)^2\,dK .
\end{aligned}
\end{equation}

\end{subequations}

\subsection{Energy distribution} \label{sec:EnergyDistr}
The theoretical framework developed above was implemented numerically in a computational toolbox, described further in Appendix \ref{A:NumericalImpl}. For a given latitude, total depth, and stratification profile, the toolbox provides a fast method to compute vertical modes, as well as the associated energy distribution for a prescribed spectral function \(S(K,j)\).
In this section, we present a series of examples illustrating the expected energy distributions obtained using our revised non-hydrostatic spectrum formulation, highlighting improvements over previous formulations.

\subsubsection{Vertical Variance}
\begin{figure*}
    \centering
    \includegraphics[width=13cm]{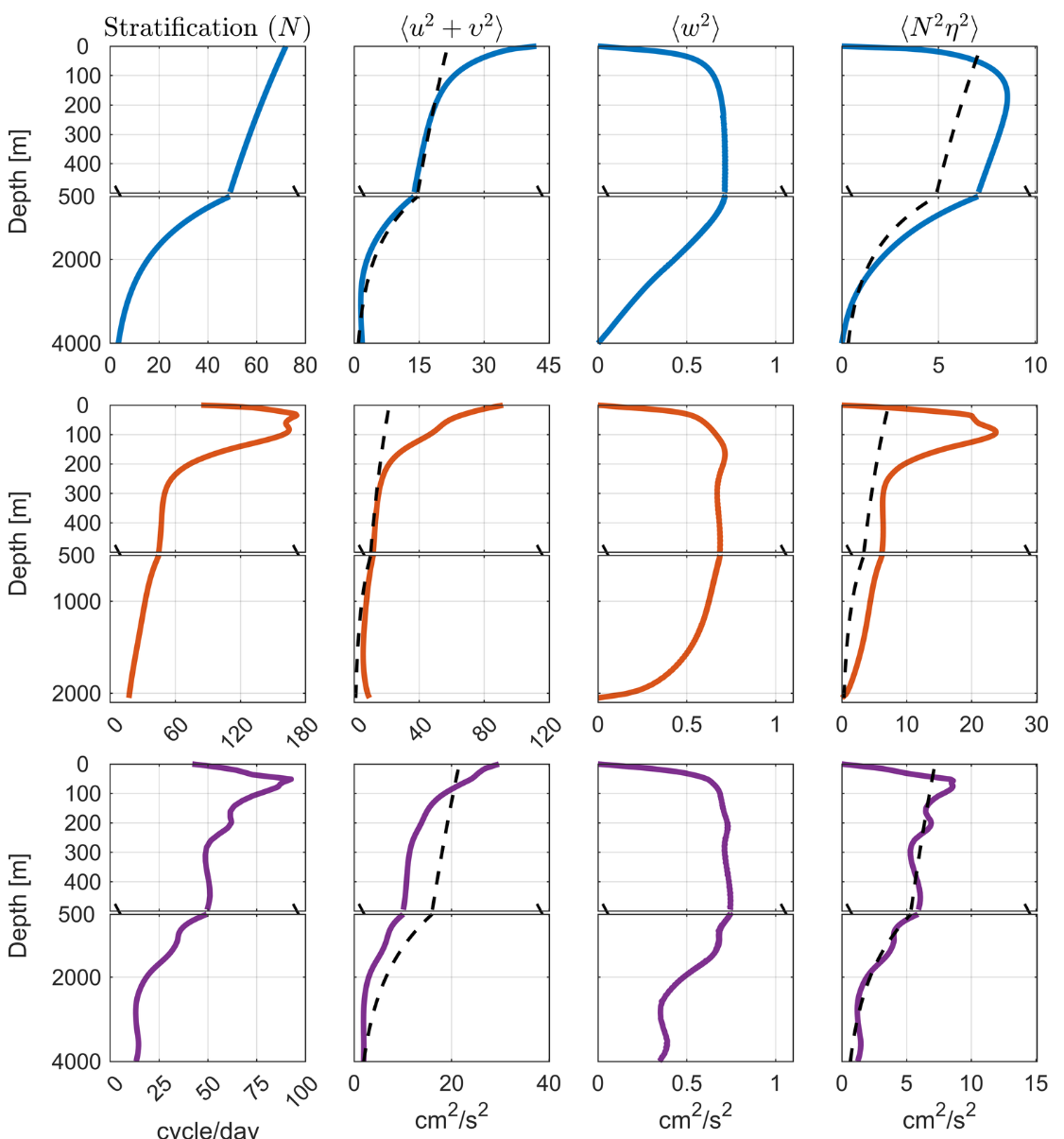}
    \caption{From left to right, each column shows vertical profiles of stratification $N(z)$, and variance of horizontal kinetic energy, vertical kinetic energy, and potential energy. Each profile is split into two depth ranges (0 to 500 m, and 500 m to local total water depth). Results are shown for three stratification profiles: the canonical Garrett–Munk (GM) exponential profile (blue), Station Papa (orange), and the Agulhas region (purple). Dashed black lines represent the expected vertical energy distribution from GM72.}
    \label{fig:verticalVariance}
\end{figure*}

Figure \ref{fig:verticalVariance} shows examples of full water column stratification profiles and associated vertical energy distributions for the non-hydrostatic formulation \eqref{eq:EK-gen} compared with the hydrostatic formulation \eqref{eq:EK-uv-eta} using the standard GM vertical structure function \eqref{vertical-structure-gm}, with the expected values of vertical displacements multiplied by $N^2(z)$ to obtain potential energy. Results for three different stratification profiles are presented: the canonical exponential stratification used in GM72, and two \textit{in situ} profiles based on mean density data obtained from the World Ocean Atlas 2018 \citep{locarnini2019}. For the \textit{in situ} profiles, the computation was performed using a single stratification profile, latitude, and local depth.

In the first example, the canonical exponential stratification (Figure \ref{fig:verticalVariance}, top-most panels), our formulation shows a clear surface intensification of HKE (top row, second panel) that is absent in GM72. This difference arises from the inclusion of vertical boundary effects in our model, which are not accounted for in GM72. Below approximately 300 m, however, both formulations produce similar HKE profiles.
By contrast, the vertical distribution of PE (top row, right-most panel) differs between the two formulations across most of the water column, a discrepancy that is, again, a direct consequence of the boundary conditions.

In the second example, from Station Papa (Figure \ref{fig:verticalVariance}, center panels), the stratification profile shows a pronounced pycnocline beneath the mixed layer, resulting in a non-monotonic $N(z)$. In this case, a wave mode may have multiple turning depths since the condition \(\omega = N(z)\) can be satisfied at different depths. As discussed in \text{\S}\ref{sec:linear-solution}\ref{Sec:TheoreticalBackground}, previous formulations fail to provide continuous solutions for non-monotonic $N(z)$, whereas the present formulation handles this without difficulty. HKE and PE profiles also show similar improvements relative to GM72 as those seen in the exponential stratification example. In the third and final example, from the Agulhas region (Figure \ref{fig:verticalVariance}, bottom-most panels), the proposed formulation again proves robust in the presence of small-scale oscillations in $N(z)$, with similar improvements relative to GM72.

Notably, the differences between the two formulations in the vertical variance are not due to non-hydrostatic effects, but to other approximations in the GM framework (see \ref{A:ApproxGM}). Because most of the internal-wave energy is in hydrostatic modes, summing over horizontal wavenumbers and vertical modes, as in Figure \ref{fig:verticalVariance}, masks the impact of the hydrostatic approximation. In the following section, we examine where the hydrostatic approximation breaks down in the frequency spectrum.

\subsubsection{Frequency spectrum}
\begin{figure}
    \centering
    \includegraphics[width=6.5cm]{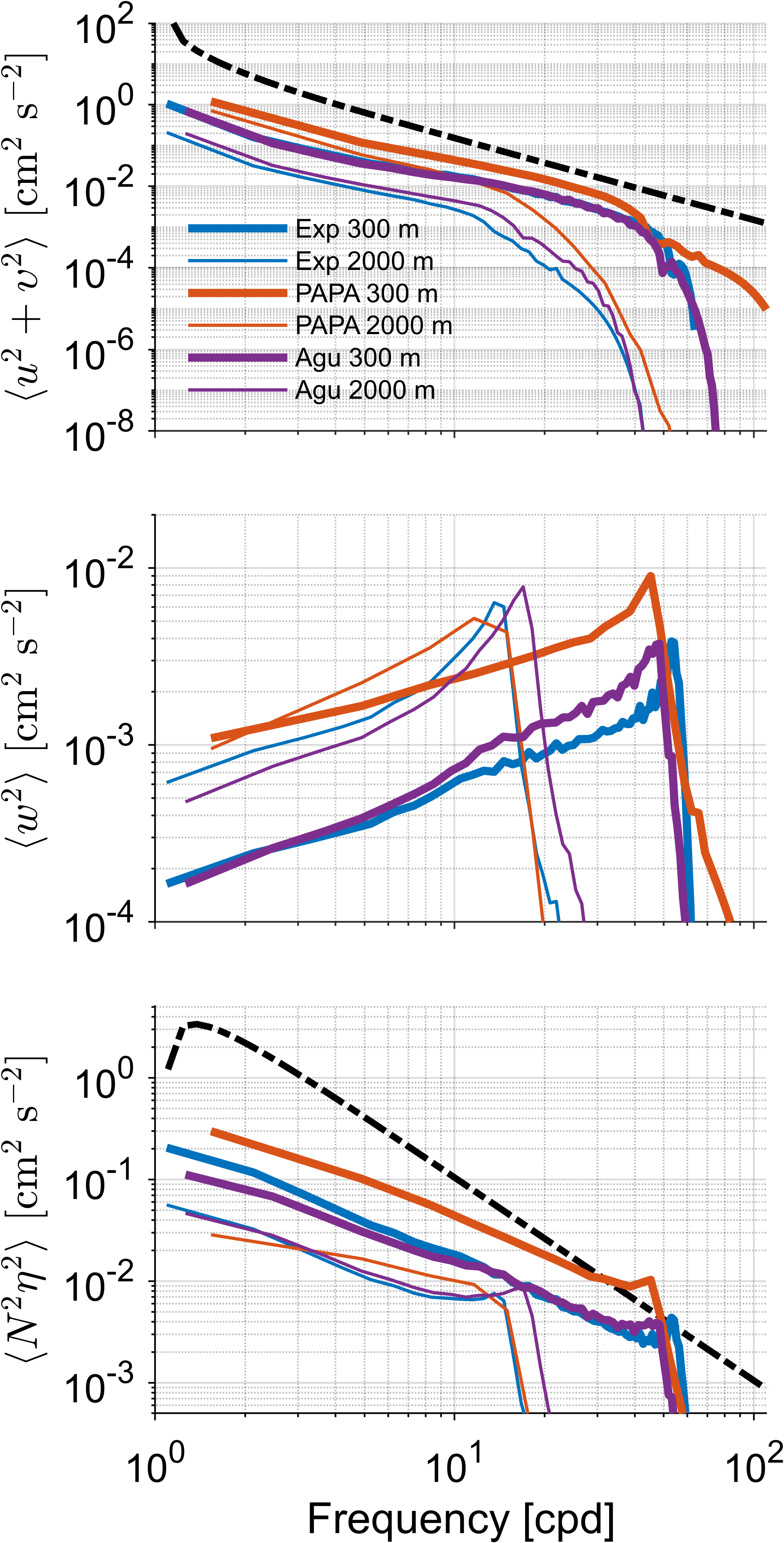}
    \caption{Frequency spectra of (top) horizontal velocity, (middle) vertical velocity, and (bottom) isopycnal displacement computed for the three stratification profiles shown in Figure \ref{fig:verticalVariance}: exponential (blue), Station Papa (orange), and Agulhas (purple). Spectra are shown at two depths: 300 m (thick line) and 2000 m (thin line).}
    \label{fig:FrequencySpectrumStations}
\end{figure}

Figure \ref{fig:FrequencySpectrumStations} shows frequency spectra for the same three stratification profiles shown in Figure \ref{fig:verticalVariance}. Note that the shape of each spectrum differs as a function of stratification profile, frequency and depth. This is in contrast to the hydrostatic GM formulation using \eqref{eq:EK-uv-eta} with \eqref{vertical-structure-gm}, where only the amplitude of the frequency spectrum depends on depth, and there is no upper bound on frequency. Such an assumption is not realistic when applied to varying ocean conditions and was one of the primary motivating factors for the formulation in \cite{Levine_2002}.

The frequency spectra of all three energy components show a clear depth dependence in both amplitude and high-frequency cut-off. This is most pronounced in VKE, which at shallower depths is relatively low at low frequencies but increases steeply toward higher frequencies, forming a pronounced peak near the local buoyancy frequency before dropping off sharply. By contrast, at greater depths spectral energy is higher at low frequencies but rolls off at a lower cut-off frequency due to the lower local buoyancy frequency.
This behavior reflects the tendency of high-frequency internal waves to be trapped near the surface, where the elevated buoyancy frequency allows them to exist.

The high-frequency peak in VKE will only appear in a statistical model if the region surrounding each turning point (associated with each frequency) is adequately represented. \citet{Desaubies_1975} demonstrated this by comparing his generalized version of GM72—which is uniformly valid in both depth and frequency (see $\S$\ref{sec:linear-solution}\ref{Sec:TheoreticalBackground})—with observed spectra from the Internal Wave Experiment (IWEX) \citep{briscoe1975_IWEX}. By contrast, the original GM72 model does not capture this peak because it excludes the singularity at the turning depth.

\subsubsection{Horizontal Wavenumber Spectrum}
The proposed spectral model, formulated in the vertical mode and horizontal wavenumber domain, enables direct computation of expected horizontal wavenumber spectra for each energy component, HKE, VKE, and PE. This provides a practical framework for characterizing the internal wavefield, as such spectra are difficult to obtain from \textit{in situ} observations, and typically require extensive spatial sampling (e.g., mooring arrays or towed ADCP systems).

\begin{figure}
    \centering
    \includegraphics[width=6.5cm]{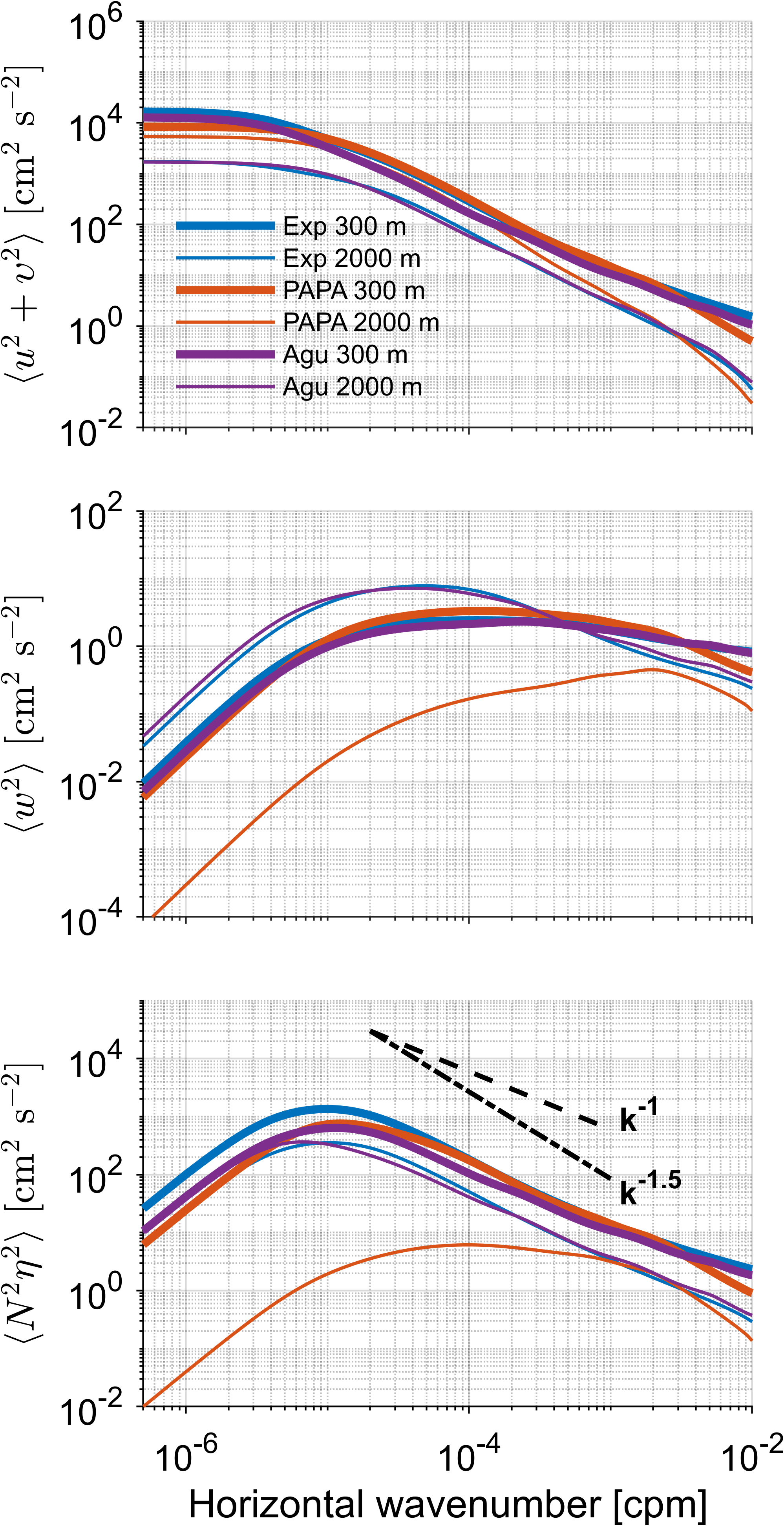}
    \caption{Horizontal wavenumber spectra of (top) horizontal velocity, (middle) vertical velocity, and (bottom) isopycnal displacement, computed for the three stratification profiles shown in Figure \ref{fig:verticalVariance}: exponential (blue), Station Papa (orange), and Agulhas (purple). Spectra are shown at two depths: 300 m (thick line) and 2000 m (thin line). Both axes are plotted on logarithmic scales.}
    \label{fig:HorizontalWavenumberSpectra}
\end{figure}

Figure \ref{fig:HorizontalWavenumberSpectra} shows horizontal wavenumber spectra of HKE, VKE, and PE for the same three stratification profiles shown in Figure \ref{fig:verticalVariance} (exponential, Station Papa, and Agulhas), again at two depths. In all three cases, HKE spectra are flat up to horizontal wavenumber $3\cdot10^{-6}$ cycle m$^{-1}$, followed by a monotonic decay with increasing wavenumber, and with higher energy levels near the surface than at depth.

By contrast, VKE and PE spectra exhibit a peak at intermediate wavenumbers, which is more pronounced for PE. This behavior is consistent with internal wave dynamics. At low horizontal wavenumbers, near-inertial waves are predominantly horizontal, resulting in relatively small VKE/HKE and PE/HKE ratios. As wavenumber increases, these ratios grow, leading to increases in VKE and PE. At higher wavenumbers, however, total energy decreases as expected for a physically realistic spectrum. The observed spectral peak therefore reflects a balance between the increasing energy partition into vertical motions and potential energy, and the overall decrease in total energy. The reduction in VKE and PE at depth is particularly evident for the Station Papa profile, where the 2000 m depth is close to the local bottom ($\approx$ 2050 m).

Because such spectra are difficult to obtain from \textit{in situ} measurements, few studies are available for comparison. \cite{katz1973} derived horizontal wavenumber spectra of isopycnal displacement from towed CTD measurements in the Sargasso Sea thermocline (550–700 m), finding an approximate $K^{-1.5}$ dependence, with a steeper, $K^{-2.3}$ decay at wavenumbers greater than 1 cycle~km$^{-1}$. \cite{Levine1986} estimated horizontal wavenumber spectra of vertical displacement from towed temperature measurements along an isobaric surface ($\approx$1184 dBar, $\approx$1200 m) near Cobb Seamount in the NE Pacific, reporting a $K^{-2}$ dependence at low wavenumbers and a transition to a $K^{-1.6}$ slope at wavenumbers above 0.03 cycle~km$^{-1}$.
Per Figure \ref{fig:HorizontalWavenumberSpectra}, the present formulation yields horizontal wavenumber spectra of vertical displacement with slopes between $K^{-1.5}$ and $K^{-1}$, roughly in agreement with previous observational findings.

\section{Discussion}
\label{sec:discussion}
In this paper we propose a set of orthogonal basis functions for non-hydrostatic ocean internal waves that allows for arbitrary, non-monotonic stratification. This framework is combined with a generalized spectral function based on GM72 as an illustrative example, though the approach can be applied to any spectral function specified in the horizontal wavenumber–vertical mode domain, including empirically derived regional spectra. While we do not claim universality for any particular spectral choice, a unified framework applicable across different oceanographic regimes represents a meaningful step toward a more flexible description of the internal wavefield.

The revised formulation addresses key limiting assumptions in GM72 and subsequent studies, including properly accounting for vertical boundary effects, the assumption that buoyancy frequency varies slowly and monotonically with depth, and that waves are hydrostatic across all frequencies.  Previous work by \cite{Levine_2002} and \cite{Desaubies_1973} attempted to address the behavior of waves near turning points, correctly expressing the expected value of the frequency spectrum of the wave field---that is, an object defined with respect to temporal Fourier modes. However, that object could not be used to directly specify a spectrum of non-hydrostatic wave modes. The present formulation resolves the orthogonality issue by formulating the spectrum in horizontal wavenumber-vertical mode space, simultaneously addressing wave mode orthogonality and allowing for non-hydrostatic waves.

Some aspects of the implementation of the theory presented here require careful consideration. First, the numerical solution of \eqref{eq:Sys} extends to horizontal wavelengths much smaller than those realized in the ocean; however, such wave solutions are physically meaningful only down to the Ozmidov scale. Second, accurate computation of vertical modes and waveguide depth requires stratification data extending to the ocean bottom. When such data are unavailable, extrapolation to the local depth is an option, though the most appropriate way to do this is an open question.

\subsection{Implications to finescale parameterization}
Applications of the GM spectrum are wide-ranging, and a revised framework for internal gravity waves has the potential to offer new perspectives, particularly for finescale parameterization of turbulent dissipation. Because direct microstructure measurements are sparse, such parameterizations infer the rate of kinetic energy dissipation ($\epsilon$) from shear and strain observations, under the assumption that internal-wave energy cascades to small scales through weakly nonlinear resonant interactions \citep{mccomas1981dynamic, Munk_1981, Gregg_1989, Polzin1995}. In this context, GM theory provides a reference for estimating spectral energy transfer in the vertical wavenumber domain \citep{Polzin_2014}. Eikonal (ray-tracing) approaches offer an alternative, describing the evolution of wave packets propagating through stochastic backgrounds modeled on the GM spectrum. These approaches assume a scale separation in which the wave packets are much smaller than the large-scale background, while allowing for non-weak nonlinear interactions \citep{Henyey_Ponphrey_1983, Henyey_Wright_Flatte_1986}. Both frameworks therefore depend, either explicitly or implicitly, on GM theory, so that in regions where the internal wave spectrum departs from the GM form, estimates of turbulent dissipation may be biased. \citet{Takahashi2021} demonstrated this explicitly, showing that finescale parameterization overestimates $\epsilon$ in the Antarctic Circumpolar Current, likely due to distortions of the vertical wavenumber spectrum relative to the GM shape, including a low-wavenumber spectral hump near 0.01 cpm. Improved representations of the internal wave spectrum, such as those developed here, may help refine the reference spectra used in finescale parameterizations, although this connection remains to be quantified.

\subsection{Importance of non-hydrostatic waves}
The role of non-hydrostatic internal waves in the ocean is not fully understood, but recent research has shed light on their significance. \cite{Pinkel_2023} demonstrated that vertical momentum transport by internal waves plays a crucial role in balancing large-scale forcing in the equatorial current system. The upper equatorial ocean experiences strong vertical shear due to the westward flow of the South Equatorial Current over the eastward Equatorial Undercurrent. Sustaining the observed current speed difference requires an unusually high effective viscosity, or vertical momentum transport. The 2012 EquatorMix Experiment revealed energetic waves with an approximate horizontal wavelength of 600 m above the Undercurrent, indicating a substantial vertical momentum flux of approximately $10^{-4}$~m$^2$s$^{-2}$.
These waves are triggered by convective cooling of the sea surface at night. They propagate downward and westward, reflecting at the critical layer, with net momentum deposition determined by the degree of dissipation during reflection.

The limited understanding of non-hydrostatic waves may partly reflect the difficulty of observing them at sufficient spatial and temporal resolution. Until recently, measurements at this scale faced significant logistical and technological constraints, including the ability to raise and lower an instrument package quickly enough, battery capacity, and the cost of densely instrumented moorings. More recent initiatives are filling this gap; the autonomous moored profiling Wirewalker, for example, is specifically designed to capture physical and biogeochemical dynamics that evolve rapidly in both time and space \citep{pinkel2011, zheng2022}. As observations increasingly access non-hydrostatic scales, internal-wave theory must also incorporate non-hydrostatic dynamics, underscoring the need for future comparisons between the framework presented here and \textit{in situ} measurements. Such comparisons will be important not only to validate theory but also to guide the design of new observational strategies.

\section{Summary and conclusions} \label{sec:conclusions}
This study presents a reformulation of internal gravity wave spectral theory that accounts for non-hydrostatic dynamics, vertical boundary effects, and arbitrary stratification. By shifting the formulation of the eigenvalue problem from the frequency domain to the horizontal wavenumber domain, energetic orthogonality among wave modes is established—an essential condition for constructing a meaningful energy spectrum. Numerical implementation, using Chebyshev polynomials and stretched vertical coordinates, accurately resolves internal wave structures even in regions of complex stratification and near turning points. The revised formulation facilitates the evaluation of internal wave energy distributions using both classical and custom spectral functions, Notably, it also captures high-frequency, surface-intensified wave energy patterns that are missed by traditional hydrostatic models. This revised framework addresses multiple limiting assumptions of the Garrett–Munk formulation and offers a practical and flexible tool for characterizing the internal wave field across a broad range of oceanographic conditions.

\appendix
\section{Orthogonality}\label{A:Othogonality}
This appendix provides the detailed derivation of the orthogonality properties summarized in \text{\S}\ref{sec:Results}\ref{Sec:Orthogonality}. We evaluate the cross terms arising from the linear operation (\ref{eq:linearOperation}) for both the fixed-$K$ and fixed-$\omega$ formulations and determine the conditions under which they vanish.
For the fixed-$K$ problem, consider two waves with distinct horizontal wavenumbers, such that either $k_a \neq k_b$ or $l_a \neq l_b$. Expressing each wave in the rhs of (\ref{eq:linearOperation}) in complex-exponential form introduces oscillatory factors in addition to the non-oscillatory terms. All cross terms have oscillatory forms that, per Fourier orthogonality, vanish when the energy is integrated horizontally over the periodic domain. More interesting is to consider the case where two waves have identical horizontal wavenumbers but different vertical modes. Evaluating the linear operation (\ref{eq:linearOperation}) the cross terms assume the form
\begin{align}
\label{eq:TEfixedTime2}
E_{\text{cross}}^{ij} &=
\frac{A^i \bar{A}^j}{4} \int_{-D}^0
\Bigg[\frac{\omega^i_K \omega^j_K + f^2}{\omega^i_K \omega^j_K}\left(F_K^i F_K^j + K^2 h_K^i h_K^j G_K^i G_K^j \right)
\\
&\qquad
+ \frac{K^2 h_K^i h_K^j}{\omega_i \omega_j}
\left(N^2 - f^2\right) G_K^i G_K^j \Bigg]\, dz + c.c. \nonumber
\end{align}
\noindent where the overbar represents the conjugate amplitude.
Substituting the orthogonality conditions (\ref{eq:K_G_eqn}) and (\ref{eq:K_F_eqn}) into (\ref{eq:TEfixedTime2}) the cross term becomes
\begin{equation}
E_{\text{cross}}^{ij}
= \frac{A^i \bar{A}^j h_K^i}{4}
\left[
\frac{\omega^i_K \omega^j_K + f^2 + K^2 g h_K^j}
{\omega^i_K \omega^j_K}
\right] \delta_{ij} \, e^{i(\theta_i-\theta_j)}.
\end{equation}
Thus $E^{ij}=0$ for $i\ne j$, demonstrating that the modes are energetically orthogonal.

The same result is not true for the case of the $\omega$-constant eigenvalue problem.
Consider two waves with same frequency, but different horizontal wavenumber and modes. Plugging them into (\ref{eq:linearOperation}), the cross terms in horizontal kinetic energy vanish, but not in vertical kinetic and potential energies---the cross term is
\begin{equation}
\begin{aligned}
E_{\text{cross}}^{ij}
&=
\frac{A^i \bar{A}^j}{2}
\int_{-D}^0\Bigg[
K^i_{\omega} K^j_{\omega} h_\omega^i h_\omega^j
+ N^2
\frac{K^i_{\omega} K^j_{\omega} h_\omega^i h_\omega^j}
{\omega^2}
\Bigg] \\
&\qquad
e^{i(\theta_i-\theta_j)}
 G_\omega^i G_\omega^j\,dz .
\end{aligned}
\end{equation}

\noindent Its not possible to use the orthogonal conditions (\ref{eq:Omega_G_eqn}) and/or (\ref{eq:Omega_F_eqn}) to eliminate all cross terms as we did for the fixed-K problem. The cross terms only vanish when the waves are hydrostatic, with no VKE, because then we can use (\ref{eq:Omega_G_eqn}) to obtain
\begin{equation}
  E_{\text{cross}}^{ij}= \frac{A^i \bar{A}^j}{2}
  \frac{K^i_{\omega} K^j_{\omega} h_\omega^i h_\omega^j}{\omega^2}
e^{i(\theta_i-\theta_j)}
g\delta_{ij}.
\end{equation}

\noindent Energy orthogonality holds for constant stratification, hydrostatic or not, because that simplifies the orthogonal conditions.

\section{The approximations of GM72}\label{A:ApproxGM}
The hydrostatic approximation made in GM72 is the topic of much of this manuscript, but there are also several other approximations made, both explicit and implicit, in the transition from the hydrostatic vertical structure functions
\begin{subequations}
\begin{align}
\label{f_structure_function}
\Phi(z) \equiv& \sum_{j=1}^{\infty} \frac{b}{h^j} H(j) (F^j(z))^2 \\
\Gamma(z) \equiv& \frac{N_0^2 b}{g} \sum_{j=1}^{\infty} H(j) (G^j(z))^2.
\end{align}
\end{subequations}
to the classic GM vertical structure functions in \eqref{vertical-structure-gm}. These approximations are 1) a WKBJ modal structure, 2) exponential-like stratification away from boundaries, and 3) a globally uniform phase speed. In net, these amount to a factor of 2 errors almost everywhere, with significantly higher errors in some parts of the water column.

\subsection{Approximation 1: WKBJ modal structure}
The WKBJ approximation to hydrostatic vertical modes (see also appendix A3 of \citealt{Early_2020}) are,
\begin{subequations}
\begin{align}
F^j_{\textrm{WKBJ}}(z) =& \sqrt{\frac{2 h^j}{L_\xi}} \sqrt{N(z)} \cos \left( \frac{j \pi \xi(z) }{L_\xi} \right) \\
G^j_{\textrm{WKBJ}}(z) =& \sqrt{\frac{2 g}{L_\xi}} \frac{1}{\sqrt{N(z)}} \sin \left( \frac{j \pi \xi(z) }{L_\xi} \right),
\end{align}
\end{subequations}
where the stretched coordinate $\xi(z)$ is defined as
\begin{equation}
\xi(z) =  \int_{-D}^z N(z^\prime) \,dz^\prime,
\end{equation}
with total length $L_\xi \equiv \xi(0)$ and eigenvalue $\sqrt{g h^j}=\frac{L_\xi}{j \pi}$. With this approximation, the dispersion relation simplifies to,
\begin{equation}
    \omega^2 = \left( \frac{L_\xi}{j \pi} \right)^2 K^2 + f^2.
\end{equation}
The WKBJ-approximated hydrostatic vertical structure functions are thus
\begin{subequations}
\label{vertical-structure-wkb}
\begin{align}
\Phi_{\textrm{WKBJ}}(z) \equiv& \frac{N(z) b}{L_\xi} \cdot 2 \sum_{j=1}^{\infty} H(j) \cos^2 \left( \frac{j \pi \xi(z) }{L_\xi} \right), \\
\Gamma_\textrm{WKBJ}(z) \equiv& \frac{N_0^2 b}{N(z) L_\xi} \cdot 2\sum_{j=1}^{\infty} H(j) \sin^2 \left( \frac{j \pi \xi(z) }{L_\xi} \right).
\end{align}
\end{subequations}
The WKBJ approximation to the hydrostatic vertical modes is remarkably accurate for exponential stratification and other smooth stratification profiles, but can still result in factor of 2 differences when using more realistic profiles, as shown in figure \ref{StructureFunctionApproximations}.

\subsection{Approximation 2: Exponential-like stratification away from boundaries}
\begin{figure}
  \centering
  \includegraphics[width=6.5cm,angle=0]{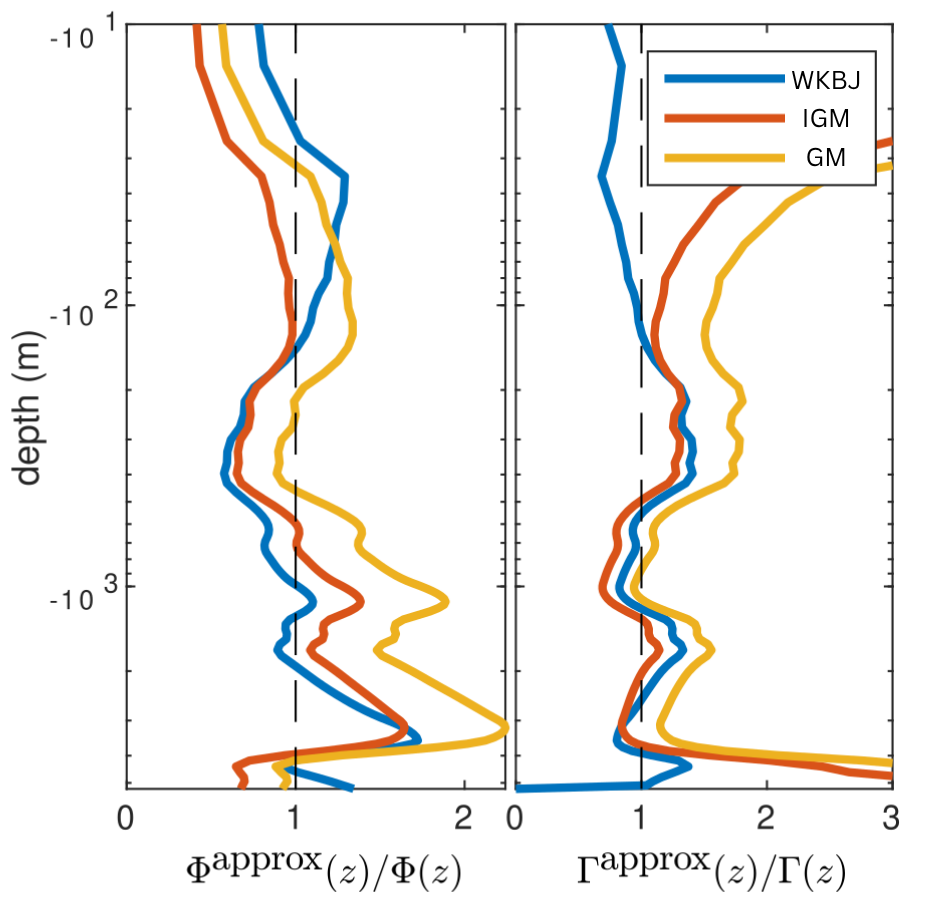}
  \caption{The three approximated versions of the vertical structure functions for the mean stratification at Woods Hole `Site D' normalized by the unapproximated version. Shown are the WKBJ approximation (blue), the intermediate GM approximation (red) and the full GM approximation (orange). Note the log scale with depth.}
  \label{StructureFunctionApproximations}
\end{figure}

The modes used by Garrett and Munk to create their empirical spectrum are simplified versions of the WKBJ solution for an exponential stratification under a very specific set of assumptions. The stratification is taken to be $N(z)=N_0 e^{z/b}$, where $b=1300$ meters and $N_0$ is 3 cycles hr$^{-1}$. With these assumptions, $L_\xi \approx N_0 b$ and the sums in \eqref{vertical-structure-wkb} are taken to be approximately $1$, i.e.,
\begin{subequations}
\begin{align}
2 \sum_{j=1}^{\infty} H(j) \cos^2 \left( j\pi e^{z/b} \right) & \approx 1\\
2 \sum_{j=1}^{\infty} H(j) \sin^2 \left( j\pi e^{z/b} \right) & \approx 1.
\end{align}
\end{subequations}
\citet{Levine_2002} plots these functions explicitly in his Figure 4. This assumption explicitly specifies the stratification in the argument of the trigonometric functions, but leaves it as unspecified outside. The result is an intermediate approximation to the Garrett-Munk vertical structure functions (IGM),
\begin{subequations}
\begin{align}
\Phi_{\textrm{IGM}}(z) =& \frac{N(z) b}{L_\xi}  \\
\Gamma_{\textrm{IGM}}(z) =& \frac{N_0^2 b}{N(z) L_\xi}.
\end{align}
\end{subequations}

This approximation is fairly severe in regions near boundaries or strong changes in stratification. Figure \ref{StructureFunctionApproximations} shows that, in some sense, this is a remarkably good approximation, although even $\Phi_{\textrm{IGM}}(z)$ shows a consistent increase in error throughout the water column. $\Gamma_{\textrm{IGM}}(z)$ in particular dramatically overestimates variance near the upper and lower boundaries by factors of 3 and 4, although it does relatively well in the interior.
The total depth-integrated energy does retain the correct GM value.
\begin{subequations}
\begin{align}
    E =& \frac{1}{2} \int_{-D}^0 (u^2 + v^2 + N^2 \eta^2) dz \\
    =& \frac{E_0}{L_\xi}  \int_{-D}^0 N(z) \, dz \\
    =& E_0
\end{align}
\end{subequations}
Finally, another limitation of this formulation is that, in practice, the value of $L_\xi$ varies significantly geographically.

\subsection{Approximation 3: Globally uniform phase speed}
One of the most severe assumptions made to the vertical structure functions is to take $L_\xi = N_0 b$, equivalent to fixing the phase speed of the fastest internal wave mode at $2.15$ m~s$^{-1}$. Figure 2 in \citet{chelton1998-jpo} shows that while this is perhaps a good typical value, there is about a factor of 2 geographic variation. With this assumption, the vertical structure functions obtain their most commonly used form,
\begin{subequations}
\begin{align}
\tag{\ref{vertical-structure-gm-phi}}
\Phi_{\textrm{GM}}(z) =& \frac{N(z)}{N_0},  \\
\tag{\ref{vertical-structure-gm-gamma}}
\Gamma_{\textrm{GM}}(z) =& \frac{N_0}{N(z)}.
\end{align}
\end{subequations}

The result of fixing the phase speed is an artificial increase or decrease in the total depth-integrated energy,
\begin{equation}
    E = E_0 \left( \frac{L_\xi}{N_0 b} \right).
\end{equation}
The error introduced by this assumption will be directly proportional to the ratio of the actual phase speed and the assumed value. For example, the actual value of the mode $j=1$ phase velocity at Woods Hole Site D is closer to 3.1 m s$^{-1}$, which is more than 1.5 times greater than the assumed GM value. Using an artificially low value causes an increased assumed variance throughout the water column, as seen by the shift to the right of the orange line in Figure \ref{StructureFunctionApproximations}.

\section{Numerical implementation}\label{A:NumericalImpl}
\label{sec:random-realizations}
In this appendix, we describe the numerical implementation of the theoretical framework developed above and its incorporation into a computational toolbox. The toolbox provides a fast method to compute vertical modes for a given latitude, total depth, and stratification profile, as well as the associated energy distribution for a prescribed spectral function \(S(K,j)\).

The numerical implementation first computes the non-hydrostatic internal wave vertical modes, \(F(z)\) and \(G(z)\). This is done by reformulating the eigenvalue problem in stretched vertical coordinates and projecting the solution onto Chebyshev polynomials. This approach accurately resolves the singular behavior that arises near turning depths. The quadrature points of the Chebyshev polynomials define an effective vertical grid, with higher resolution in regions where vertical and temporal scales are small, such as near a pycnocline. Further details on the numerical computation of vertical modes can be found in \cite{Early_2020}.

Once the vertical modes are known, the energy coefficients associated with each component of internal wave energy are computed. The next step is to define a prescribed spectral function, \(S(K,j)\). By default, the toolbox uses \eqref{eq:generalSpec}, though a custom function, such as one derived from a region specific characterization of the internal wave field, may also be provided. The spectral function is normalized such as $\int_K \sum_j S(K,j)= 1$, and is integrated over wavenumber bands to compute the total energy in each mode and wavenumber bin (Algorithm \ref{alg:amplitudesWithSpectrum}). Multiplying the total energy in each bin by the corresponding energy coefficients and vertical modes produces estimates of the energy components, HKE, VKE, and PE, as functions of
($K,j,z$)\ (Algorithm \ref{alg:assignEnergy}).

As introduced in \text{\S}\ref{sec:linear-solution}\ref{subsec:ensemble}, the internal-wave field is modeled as an ensemble of linear waves whose amplitudes are treated as independent zero-mean Gaussian random variables, allowing observable quantities to be described through expectation values. Next, we summarize how the expected statistical properties arise and how a stochastic realization from a prescribed internal-wave energy spectrum was implemented in the toolbox (Algorithm \ref{alg:randomSpectrum}).

Each \((K,j)\) band contains \(n\) statistically independent waves. Because the spectral model only provides the total band energy \(E_{K,j}\), the energy is distributed uniformly among the waves, giving each wave an energy \(\varepsilon_{k,l,j}=E_{K,j}/n\). The classical assumption is that internal wave phases and amplitudes are random, so we assume the amplitudes to be normally distributed. The normalized amplitude of a single wave is written as
\begin{equation}
    A_{k,l,j} = X_{k,l,j}\sqrt{2\varepsilon_{k,l,j}/h^j_K},
\end{equation}

\noindent where \(X_{k,l,j}\) is a normally distributed random variable with zero mean and unit variance. This choice ensures that the amplitude variance is consistent with the specified energy per wave \(\varepsilon_{k,l,j}\). The total squared amplitude in the \((K,j)\) band then is obtained by summing the contributions from the \(n\) waves in the band, as
\begin{equation}
A_{K,j}^2=\sum_{l=1}^n A_{k,l,j}^2
= \frac{2E_{K,j}}{h^j_Kn}\sum_{l=1}^n X_{k,l,j}^2 =\frac{2E_{K,j}}{h^j_Kn} Z,
\end{equation}

 \noindent where \(\sum_{l=1}^n X_{k,l,j}^2\) follows a chi-squared distribution with \(n\) degrees of freedom, \(Z\sim\chi^2(n)\). The mean and variance of $Z$ are, respectively, \(n\) and \(2n\). Therefore mean and variance of the squared amplitude distribution in each band is given, respectively, by
 \begin{subequations}
     \begin{align}
         \label{eq:chiMean}
         \langle A_{K,j}^2 \rangle &= 2E_{K,j}/h^j_K\\
         \label{eq:chiVar}
         \mathrm{var}\langle A_{K,j}^2 \rangle &= 8nE_{K,j}^2/([h^j_K]^2).
     \end{align}
 \end{subequations}
Thus, as \(n\) increases, the variance decreases, meaning that bands populated by many waves become increasingly deterministic. This captures the usual assumption in spectral internal-wave models that sufficiently dense sampling in wavenumber space allows stochastic realizations to reproduce the mean spectral shape with negligible variance.

\begin{algorithm}
\caption{Computation of $K$-band-integrated energy from a prescribed spectrum}
\label{alg:amplitudesWithSpectrum}
\begin{algorithmic}[1]

\Require Discrete radial wavenumber grid $\{K_i\}$, vertical modes $j$, spectral function $S(K,j)$
\Ensure Band-integrated energy $E_{K,j}$

\State Initialize $E_{K,j} = 0$

\State Define bin edges:
\[
K_{i+\frac{1}{2}} = \sqrt{K_i K_{i+1}}
\]
\For{each vertical mode $j$}
    \For{each wavenumber bin $K_i$}
        \State Compute band energy:
        \[
        E_{K_i,j} = \int_{K_{i-\frac{1}{2}}}^{K_{i+\frac{1}{2}}} S(K,j)\, dK
        \]
        \State Compute spectral density:
        \[
        S_{K_i,j} = \frac{E_{K_i,j}}{K_{i+\frac{1}{2}} - K_{i-\frac{1}{2}}}
        \]
    \EndFor
\EndFor
\If{verbose}
    \For{each mode $j$}
        \State Compare $\int S(K,j)\,dK$ with $\sum_i E_{K_i,j}$
    \EndFor
\EndIf
\end{algorithmic}
\end{algorithm}

\begin{algorithm}
\caption{Assignment of wave-energy components from a prescribed spectrum}
\label{alg:assignEnergy}
\begin{algorithmic}[1]

\Require Model object containing modal structure and coefficients; prescribed spectral function $S(K,j)$
\Ensure Updated model object with total energy and energy components
\State Validate the spectral function by evaluating it at a test point
\If{$S(K,j)$ cannot be evaluated}
    \State Stop and return an error
\EndIf
\State Compute the band-integrated total energy:
\[
E_{K,j} \gets \textsc{amplitudesWithSpectrum}(S) \quad \text{(see Algorithm~\ref{alg:amplitudesWithSpectrum})}
\]

\State Compute the squared wave amplitudes:
\[
A_{K,j}^2 = \frac{2E_{K,j}}{h_K^j}
\]

\State Evaluate the buoyancy frequency profile at the modal quadrature points:
\[
N^2(z_q) \gets N^2(z_q)
\]

\State Compute horizontal kinetic energy:
\[
\mathrm{HKE}(K,j,z) = A_{K,j}^2 \, C_{\mathrm{HKE}}(K,j)\, F^2(z)
\]

\State Compute vertical kinetic energy:
\[
\mathrm{VKE}(K,j,z) = A_{K,j}^2 \, C_{\mathrm{VKE}}(K,j)\, G^2(z)
\]

\If{$N^2$ is constant}
    \State Use the scalar value of $N^2$
\Else
    \State Reshape $N^2$ to conform with the dimensions of $(K,j,z)$
\EndIf

\State Compute potential energy:
\[
\mathrm{PE}(K,j,z) = A_{K,j}^2 \, C_{\mathrm{PE}}(K,j)\, G^2(z)\, N^2(z)
\]

\State Store $A_{K,j}^2$, $E_{K,j}$, HKE, VKE, PE, and $N^2(z_q)$ in the model object

\State \Return updated model object

\end{algorithmic}
\end{algorithm}
\begin{algorithm}
\caption{Generation of a stochastic realization of the internal-wave spectrum}
\label{alg:randomSpectrum}
\begin{algorithmic}[1]

\Require Variance field $A^2_{K,j}$ computed from Algorithm~\ref{alg:assignEnergy}, radial wavenumber grid $\{K_i\}$, eigendepth $h_K^j$
\Ensure Random realization of the spectrum $S^{\mathrm{rnd}}_{K,j}$

\State Define the number of samples per wavenumber bin:
\[
n(K_i) \propto \frac{K_i}{K_{\min}}
\]

\For{each vertical mode $j$}
    \For{each wavenumber bin $K_i$}

        \State Draw $n(K_i)$ independent Gaussian samples:
        \[
        X_l \sim \mathcal{N}\left(0,\, A^2_{K_i,j}\right)
        \]

        \State Form squared amplitudes:
        \[
        X_l^2
        \]

        \State Compute sample mean:
        \[
        \widetilde{A^2}_{K_i,j} = \frac{1}{n(K_i)} \sum_{l=1}^{n(K_i)} X_l^2
        \]

    \EndFor
\EndFor

\State Convert variance to energy:
\[
S^{\mathrm{rnd}}_{K,j} = \frac{h_K^j}{2} \, \widetilde{A^2}_{K,j}
\]

\State \Return $S^{\mathrm{rnd}}_{K,j}$

\end{algorithmic}
\end{algorithm}

\clearpage

\acknowledgments
We thank Gerardo Hernández-Dueñas for checking the orthogonality proof.
The authors gratefully acknowledge the financial support provided by the National Science Foundation under Grant Nos. 2123394 and 2123740.

\datastatement
The analysis scripts used in this study are publicly available at
\url{https://doi.org/10.5281/zenodo.20148657}. The Internal Gravity
Wave Spectrum toolbox is available at
\url{https://doi.org/10.5281/zenodo.20148339}.

 \bibliographystyle{ametsocV6}
 \bibliography{references}

\end{document}